\documentclass[aip,jcp,reprint,nobibnotes,superscriptaddress,floatfix, 
showpacs,citeautoscript,showkeys]{revtex4-1}

\usepackage{graphicx}
\usepackage{scalerel}
\usepackage[version=4]{mhchem}
\usepackage{latexsym}
\usepackage{textcomp}
\usepackage{amssymb}
\usepackage{amsbsy}
\usepackage{amsmath}
\usepackage{epstopdf}
\usepackage{bm}
\usepackage{longtable}
\usepackage{subfigure}
\usepackage{epsfig}
\usepackage{dcolumn}
\usepackage[Euler]{upgreek}
\usepackage{multirow}
\usepackage{footnote}
\usepackage{stmaryrd}
\usepackage{tabulary}
\usepackage{color}
\usepackage{xcolor}
\usepackage{tikz}
\usetikzlibrary{shapes}
\usepackage{mathtools}
\usepackage{empheq}
\usepackage{cancel}

\usepackage{breqn}

\makeatletter
\let\cat@comma@active\@empty
\makeatother

\begin{document}
\title{Universality of the collapse transition of sticky polymers}
\date{\today}
\author{Aritra Santra}
\affiliation{Department of Chemical Engineering, Monash University,
Melbourne, VIC 3800, Australia}
\author{Kiran Kumari}
\affiliation{IITB-Monash Research Academy, Indian Institute of Technology Bombay, Mumbai, Maharashtra -  400076, India}
\affiliation{Department of Biosciences and Bioengineering, Indian Institute of Technology Bombay, Mumbai, Maharashtra -  400076, India}
\affiliation{Department of Chemical Engineering, Monash University,
Melbourne, VIC 3800, Australia}
\author{R. Padinhateeri}
\affiliation{Department of Biosciences and Bioengineering, Indian Institute of Technology Bombay, Mumbai, Maharashtra -  400076, India}
\author{B. D\"{u}nweg}
\affiliation{Max Plank Institute of Polymer Research, Ackermannweg 10, 55128 Mainz, Germany}
\affiliation{Department of Chemical Engineering, Monash University,
Melbourne, VIC 3800, Australia}
\author{J. Ravi Prakash}
\email[Electronic mail: ]{ravi.jagadeeshan@monash.edu}
\affiliation{Department of Chemical Engineering, Monash University,
Melbourne, VIC 3800, Australia}


\begin{abstract}
The universality of the swelling of the radius of gyration of a homopolymer relative to its value in the $\theta$ state, independent of polymer-solvent chemistry, in the crossover regime between $\theta$ and athermal solvent conditions, is well known. Here we study, by Brownian dynamics, a polymer model where a subset of monomers is labelled as ``stickers''. The mutual interaction of the stickers is more attractive than those of the other (``backbone") monomers, and has the additional important characteristic of ``functionality" $\varphi$, i.e., the maximum number of stickers that can locally bind to a given sticker. A saturated bond formed in this manner remains bound until it breaks due to thermal fluctuations, a requirement which can be viewed as an additional Boolean degree of freedom that describes the bonding. This, in turn, makes the question of the order of the collapse transition a non-trivial one. Nevertheless, for the parameters that we have studied (in particular, $\varphi=1$), we find a standard second-order $\theta$ collapse, using a renormalised solvent quality parameter that takes into account the increased average attraction due to the presence of stickers. We examine the swelling of the radius of gyration of such a sticky polymer relative to its value in the altered $\theta$ state, using a novel potential to model the various excluded volume interactions that occur between the monomers on the chain. We find that the swelling of such sticky polymers is identical to the universal swelling of homopolymers in the thermal crossover regime. Additionally, for our model, the Kuhn segment length under $\theta$ conditions is found to be the same for chains with and without stickers. 
\end{abstract}

 
\maketitle

\section{\label{sec:intro}Introduction}

Solutions of sticky polymers consist of chains with sticky groups that can form reversible physical bonds, which in turn lead to the formation of reversible gels and networks. The ability to tune different microscopic parameters of sticky polymer chains, like the number of stickers per chain, the position of the stickers on the polymer backbone, the strength of associations of the stickers and the solution temperature (or backbone solvent quality) has led to the use of sticky polymer solutions in a number of different applications, such as rheology modifiers, adhesives, biomedical implants, adsorbents and many  such more\cite{RheoMod,BioPharm}. Due to the relative affinity of sticky groups for each other, sticky polymer chains are more collapsed or less swollen at a given temperature compared to the corresponding homopolymer made up of only the backbone or non-sticky monomers of the same molecular weight. For example, at the $\theta$ temperature for the homopolymer, simple linear polymer chains in a dilute solution follow random walk (RW) statistics, whereas, the introduction of sticky groups leads to a decrease in the size of the chain due to relatively poorer solvent quality. Indeed the whole phase diagram for homopolymers~\cite{Grosberg, RubCol} is expected to be modified due to the presence of stickers~\cite{RnSstatics,RnSdynamics,Dob}. In many applications like mist control or drag reduction of aviation fuel, it is necessary to have long, swollen, physically associated polymer chains in a single-phase solution\cite{Kornfield1,Kornfield2}. For such applications, it is important to estimate the renormalised solvent quality of a sticky polymer solution in order to have some knowledge of chain conformations and the relative location of the system in the phase-space of temperature and concentration. In this paper we address the question of how to compute the solvent quality of dilute sticky polymer solutions and show that the swelling behaviour of sticky polymers in terms of a renormalised solvent quality follows the same universal behaviour as a standard second-order $\theta$ transition. As will be discussed subsequently, we consider the possibility that a first-order collapse might also exist in a certain parameter region for the particular model studied here, making the current observation a non-trivial one. 

For homopolymers, it is well-known that in the limit of large molecular weight, static properties of polymer chains in dilute solution, such as the radius of gyration $R_g$, follow universal power laws in both $\theta$ and athermal solvents. Furthermore, experiments and theoretical studies indicate that in the region between $\theta$ and athermal solvents universal behaviour in terms of crossover scaling is still observed. For such systems the mean size of the polymer is a function of both the temperature ($T$) and the molecular weight ($M$) which combine to form a single variable, the solvent quality, $z = k(1-T_{\theta}/T)\sqrt{M}$, where $T_{\theta}$ indicates the temperature corresponding to a $\theta$-solvent and $k$ is a chemistry-dependent constant. A plot of the swelling ratio, $\alpha_g$, which is the ratio of $R_g$ in a good solvent to that in a $\theta$-solvent, against the solvent quality $z$, for $T>T_{\theta}$, collapses data on a universal master curve for a wide variety of polymer-solvent systems with an appropriate choice of the constant $k$\cite{Hayward,Fujita,Rondelez,Ioan}. Within the framework of Brownian dynamics (BD) simulations, this collapse has been demonstrated by Kumar and Prakash~\cite{KumarRavi}. 

To the best of our knowledge, there are no studies exploring the universal swelling behaviour of dilute solutions of sticky polymers. In this paper we have used a novel potential, proposed by Soddemann et al.\cite{SDK} (which  we denote as the SDK potential) to investigate the effect of stickers on the solvent quality and the swelling behaviour of sticky polymers in dilute solutions, using the methodology of Kumar and Prakash~\cite{KumarRavi}. The sticky macromolecules are modelled as multi-sticker chains with $f$ equispaced stickers positioned along the backbone of each chain (except at the chain ends where there are no stickers) separated by $\ell$ spacer (or backbone) monomers. A sticker is assumed to associate with only one other sticker (i.e. with functionality $\varphi =1$). Such systems can be easily designed in experiments\cite{Kornfield1,Kornfield2,Guo2005}. The key idea is to investigate various systems characterised by different parameters, and to check if the typical characteristics of a second-order $\theta$ collapse are observed. Since this indeed appears to be valid in our model, we are able to verify the universality of the swelling of sticky polymer solutions. 

The advantage of using the SDK potential is that it can be used to represent both the backbone monomer-monomer interactions, and the sticker monomer-monomer interactions, with a simple choice of the attractive well-depth of the potential. In the former case this is denoted by $\epsilon_{bb}$, while in the latter it is denoted by $\epsilon_{st}$. We find that the effective solvent quality of a sticky polymer solution can be represented in terms of these variables, along with the clear identification of the $\theta$-temperature. As a consequence, the swelling of sticky polymer chains can be examined as a function of the various parameters that control their static properties.

The plan of the paper is as follows. Section~\ref{sec:SM} discusses the principal governing equations and the details of the various interactions.  The description of the universal swelling of homopolymers, which forms the framework within which the universal behaviour of sticky polymer solutions is discussed, is taken up in Section~\ref{sec:UniSwellHP}. The determination of the $\theta$-temperature of sticky polymer solutions, in terms of the appropriate value of the well depth of the SDK potential, is considered in~\ref{sec:thetaCalc}, while the universal swelling of sticky polymers is examined in~\ref{sec:UniswellSP}. The main  conclusions are summarised in Section~\ref{conclusions}. In order to make the discussion of sticky polymer solutions the central focus of the paper, the estimation of the well depth of the SDK potential that determines the $\theta$-point for homopolymer solutions is described in an Appendix, while the optimisation of the cut-off radius of the SDK potential and the considerations that lead to the choice of a particular value, are discussed in the Supplementary Information. 

\section{\label{sec:SM} Basic equations and simulation details}

On the mesoscopic scale, polymers are modelled as a sequence of coarse-grained bead-spring chains with $N_b$ beads connected by $N_b-1$ entropic springs\cite{Bird}. In this study we have simulated a single chain, in an implicit-solvent framework, with the chain configuration specified by the set of position vectors $\mathbf{r}_\mu (\mu = 1, 2, ..., N_b)$. The evolution of bead positions in BD simulations is governed by the following Ito stochastic differential equation,
\begin{widetext}
\begin{align}\label{gov-eqn}
\begin{aligned}
\mathbf{r}_\mu(t + \Delta t) =\, & \mathbf{r}_\mu(t) + \frac{\Delta t}{4} \sum\limits_{\nu=1}^N\mathbf D_{\mu\nu}\cdot(\mathbf F_\nu^{s}+ \mathbf F_\nu^{\textrm{SDK}}) +\frac{1}{\sqrt{2}}\sum\limits_{\nu=1}^N \mathbf B_{\mu\nu}\cdot\Delta \mathbf W_\nu
\end{aligned}
\end{align}
\end{widetext}
Here the length and time scales are non-dimensionalised with $l_H=\sqrt{k_BT/H}$ and $\lambda_H=\zeta/4H$, respectively where $T$ is the absolute temperature, $k_B$ is the Boltzmann constant, $H$ is the spring constant, and $\zeta=6\pi\eta_s a$ is the Stokes friction coefficient of a spherical bead of radius $a$ where $\eta_s$ is the solvent viscosity. $\Delta\pmb W_\nu$ is a non-dimensional Wiener process with mean zero and variance $\Delta t$. $\pmb{B}_{\mu\nu }$ is a non-dimensional tensor whose evaluation requires the decomposition of the diffusion tensor $\pmb D_{\nu\mu}$, defined as $\pmb D_{\nu\mu} = \delta_{\mu\nu} \pmb \delta + \pmb \Omega_{\mu\nu}$, where $\delta_{\mu\nu}$ is the Kronecker delta, $\pmb \delta$ is the unit tensor, and $\pmb{\Omega}_{\mu\nu}$ is the hydrodynamic interaction tensor. Defining the matrices $\mathcal{D}$ and $\mathcal{B}$ as block matrices consisting of $N \times N$ blocks each having dimensions of $3 \times 3$, with the $(\mu,\nu)$-th block of $\mathcal{D}$ containing the components of the diffusion tensor $\pmb{D}_{\mu\nu }$, and the corresponding block of $\mathcal{B}$ being equal to $\pmb{B}_{ \mu\nu}$, the decomposition rule for obtaining $\mathcal{B}$ can be expressed as $\mathcal{B} \cdot {\mathcal{B}}^{\textsc{t}} = \mathcal{D}\label{decomp}$. The bonded interactions between the beads are represented by a spring force, $\mathbf F_\nu^{s}$, and the non-bonded excluded volume (EV) interactions are denoted by $\mathbf F_\nu^{\textrm{SDK}}$. We use the regularized Rotne-Prager-Yamakawa (RPY) tensor to compute hydrodynamic interactions (HI),
\begin{equation}
{\pmb{\Omega}_{\mu \nu}} = {\pmb{\Omega}} ( {\mathbf{r}_{\mu}} - {\mathbf{r}_{\nu}} )
\end{equation}
where 
\begin{equation}
\pmb{\Omega}(\mathbf{r}) =  {\Omega_1{ \pmb \delta} +\Omega_2\frac{\mathbf{r r}}{{r}^2}}
\end{equation}
with
\begin{equation*}
\Omega_1 = \begin{cases} \dfrac{3\sqrt{\pi}}{4} \dfrac{h^*}{r}\left({1+\dfrac{2\pi}{3}\dfrac{{h^*}^2}{{r}^2}}\right) & \text{for} \quad r\ge2\sqrt{\pi}h^* \\
 1- \dfrac{9}{32} \dfrac{r}{h^*\sqrt{\pi}} & \text{for} \quad r\leq 2\sqrt{\pi}h^* 
\end{cases}
\end{equation*}
and 
\begin{equation*}
\Omega_2 = \begin{cases} \dfrac{3\sqrt{\pi}}{4} \dfrac{h^*}{r} \left({1-\dfrac{2\pi}{3}\dfrac{{h^*}^2}{{r}^2}}\right) & \text{for} \quad r\ge2\sqrt{\pi}h^* \\
 \dfrac{3}{32} \dfrac{r}{h^*\sqrt{\pi}} & \text{for} \quad r\leq 2\sqrt{\pi}h^* 
\end{cases}
\end{equation*}
Here, the hydrodynamic interaction parameter $h^*$ is the dimensionless bead radius in the bead-spring model, defined as $h^* = a/(\sqrt{\pi k_BT/H})$. In all the simulations reported here in which hydrodynamic interactions have been implemented, a value of $h^{\ast}$ equal to 0.25 has been used. Unless explicitly stated, however, hydrodynamic interactions have been turned off by setting ${\pmb{\Omega}_{\mu \nu}} =\mathbf{0}$.

We use a finitely extensible nonlinear elastic (FENE) spring potential to represent the interaction between adjacent beads,
\begin{align}\label{eq-fraenkel}
U_{\textrm{FENE}}= -\frac{1}{2}Q_0^2\ln \left( 1-\frac{r^{2}}{Q_0^2}\right)
\end{align}
where $Q_0$ is the dimensionless maximum stretchable length of a single spring, and $k_B T$ is used to non-dimensionalise energy. All the simulations reported in this work use a value of $Q_0^2 = 50.0$. Note that the notation $Q_0^2$ used here is the same as the more commonly used FENE $b$-parameter.  The spring force on a bead resulting from $U_{\textrm{FENE}}$ is denoted by $\mathbf{F}_\nu^{s}$. 

Note that a large value of $Q_0$, as used here, implies a very soft potential that admits the possibility of self-crossing of the chain. This would pose a severe problem if we were interested in the \emph{dynamics of dense systems}. However, our present investigation aims at statics in the dilute limit, and future studies on dynamics will be restricted to the dilute limit too. It is well known that for these properties topological constraints do not play a role. Rather on the contrary, self-crossings are expected to speed up the exploration of phase space, and are hence advantageous for our purposes. 

The excluded volume interactions between pairs of beads on the chain is modelled by a novel potential, $U_{\text{SDK}}$, proposed by Soddemann-D\"{u}nweg-Kremer~\cite{SDK},
\begin{align}\label{eq:SDK}
U_{\textrm{SDK}}=\left\{
\begin{array}{l l l}
&4\left[ \left( \dfrac{\sigma}{r} \right)^{12} - \left( \dfrac{\sigma}{r} \right)^6 + \dfrac{1}{4} \right] - \epsilon;  & r\leq 2^{1/6}\sigma \vspace{0.5cm} \\
& \dfrac{1}{2} \epsilon \left[ \cos \,(\alpha \left(\dfrac{r}{\sigma}\right)^2+ \beta) - 1 \right] ;& 2^{1/6}\sigma \leq r \leq r_c \vspace{0.5cm} \\
& 0; &  r \geq r_c
\end{array}\right.
\end{align}
The potential has a minimum at $r=2^{1/6}\sigma$, and the value of the non-dimensional distance $\sigma$ is taken to be 1 in the present study. The quantity $\epsilon$ is the attractive well depth of the potential. As can be seen from Eq.~(\ref{eq:SDK}), the repulsive part of the SDK potential is modelled by a truncated Lennard-Jones (LJ) potential similar to the Weeks-Chandler-Anderson (WCA) potential while the attractive contribution is modelled with a cosine function. Unlike the LJ potential, which has a long attractive tail, the short ranged attractive tail of the SDK potential smoothly approaches zero at a finite distance $r_c$, which leads to an increase in the simulation efficiency\cite{SDK}. It is worth noting that $\epsilon = 0$ in the SDK potential corresponds to a  purely repulsive WCA potential, and the solvent quality reduces with increasing values of $\epsilon$. An advantage of the SDK potential over the LJ potential is that the complete range of solvent qualities, from poor to athermal, can be explored by varying the single parameter, $\epsilon$, which can alter the attractive component of the SDK potential without affecting the repulsive force. The constants $\alpha$ and $\beta$ are determined by applying the two boundary conditions, namely, $U_{\text{SDK}} = 0$ at $r=r_c$, and $U_{\text{SDK}}=-\epsilon$ at $r=2^{1/6}\sigma$. Based on these two boundary conditions, $\alpha$ and $\beta$ are calculated by solving the following set of equations,
\begin{align}
\label{alphabeta}
2^{1/3}\alpha + \beta &=\pi \\
\left(\frac{r_c}{\sigma}\right)^2 \alpha + \beta &= 2\pi 
\end{align} 
In order to solve the above set of equations, it is required to choose a reasonable value of the cut-off radius, $r_c$. In the original study by Soddemann et al.\cite{SDK} the cut-off radius of the potential, $r_c$,  was chosen to be $1.5\,\sigma$ in order to include only the first neighboring shell of interactions, determined from the first minimum of the pair correlation function. For $r_c=1.5\,\sigma$, the values of $\alpha$ and $\beta$ are calculated to be $3.1730728678$ and $-0.856228645$, respectively\cite{SDK}, and the resultant SDK potential has been used to investigate various equilibrium properties of the solutions of polymer chains using molecular dynamics (MD) and Monte Carlo (MC) simulations\cite{SDK,Steinhauser}. In the context of the current Brownian dynamics simulations, however, we find that using $r_c=1.5\,\sigma$ leads to the prediction of unphysical asymptotic scaling behaviour in the poor solvent limit. We were able to ``cure'' the problem by using a value of $r_c = 1.82 \, \sigma$. A detailed discussion of the problems encountered with the original cut-off radius, and the process by which the revised value was arrived at is given in the Supplementary Information. All subsequent results reported here with the SDK potential are for $r_c = 1.82 \,\sigma$. 

At this point, it is appropriate to elaborate on what is meant by the concept of the functionality $\varphi$, and the choice of its value of one in the model. For each pair of monomers $\mu$ and $\nu$, we introduce a Boolean variable $q_{\mu\nu} \in \{0,1\}$. $q_{\mu\nu}$ is zero whenever at least one of the two monomers is a backbone monomer, while for a pair of sticker monomers, $q_{\mu\nu}$ is zero if no bond exists between the two stickers, and $q_{\mu\nu} = 1$ for a bonded pair of sticker monomers. The attractive strength, $\epsilon$ of the SDK potential for a pair of monomers $\mu$ and $\nu$ is then given by
\begin{equation} 
\label{Boolean}
\epsilon = (1- q_{\mu\nu})\, \epsilon_{bb} + q_{\mu\nu} \,\epsilon_{st} 
\end{equation} 
where, as mentioned earlier, backbone monomer-monomer interactions are denoted by $\epsilon_{bb}$, and the sticker monomer-monomer interactions are denoted by $\epsilon_{st}$. Typically, $\epsilon_{st} \ge \epsilon_{bb}$. The variables $q_{\mu\nu}$ can be considered to be additional degrees of freedom whose dynamics are coupled to the dynamics of the monomer coordinates according to the following simple update rules:
\begin{enumerate}
\item Whenever two stickers come within the cutoff radius of the SDK potential, $r_c$, the value of $q_{\mu\nu}$ is changed from zero to one, provided that both stickers are not bonded to other stickers.
\item As long as these two stickers are within the interaction range $r_c$, $q_{\mu\nu}$ is maintained at the value one, regardless of how closely they might be approached by other stickers. 
\item As soon as the distance between two stickers becomes greater than $r_c$, $q_{\mu\nu}$ is reset to zero, and new bondings may occur. 
\end{enumerate}
It is clear that the monomer coordinates, together with the $q_{\mu\nu}$ values, provide sufficient information to calculate the interaction energy of the system uniquely. Furthermore, it is also clear that the update rules give rise to a well defined configuration space of the system in the sense of Statistical Mechanics, such that the partition function exists. As long as there are only pairs of stickers $\mu$ and $\nu$ within interaction range, it is clear that the corresponding $q_{\mu\nu}$ has to be one, while in a situation, where, for example, three stickers are all within interaction range of each other, there are three possibilities to form the bond, corresponding to the three sides of the triangle formed by the three stickers. Since all three cases are dynamically accessible, each of them must appear in the partition function. In this context, it should be emphasised that our simulation setup aims at modelling \emph{reversible} sticker bonds. 

Note also that other update rules, and/or other values of $\varphi$, may well be conceivable, which would then give rise to a different configuration space, and a correspondingly altered Statistical Mechanics of the system. 

The observation that we are dealing with additional degrees of freedom makes the existence of a second-order $\theta$ transition a non-trivial and subtle question. There are various examples in Statistical Mechanics where the coupling to an additional degree of freedom turns a second-order phase transition into a first-order transition. Some of these examples, which are relevant to the present work, are discussed in Section~\ref{conclusions}.

The simulations are carried out for different chain lengths, $N_b$, ranging from 25 to 90 beads per chain, with an equilibration run of about $8$ Rouse relaxation times (estimated analytically as given in~\citet{Bird}) and a production run of $6$ to $8$ Rouse relaxation times  with a non-dimensional time step size $\Delta t =0.001$. Data from each independent trajectory in the simulations are collected at an interval of $1000$ to $5000$ non-dimensional time steps over the entire production run and time averages are calculated over each of the trajectories. Average equilibrium properties and error of mean estimates are evaluated over an ensemble of such independent time averages consisting of $1000$ to $2000$ independent trajectories. In the case of sticky polymers, an additional pre-equilibration run of 2 to 3 Rouse times is carried out with a chain without stickers.

\section{\label{sec:UniSwellHP} Universal swelling of homopolymers}

\begin{figure}[tbh]
	\begin{center}
	 {\includegraphics[width=3.5in,height=!]{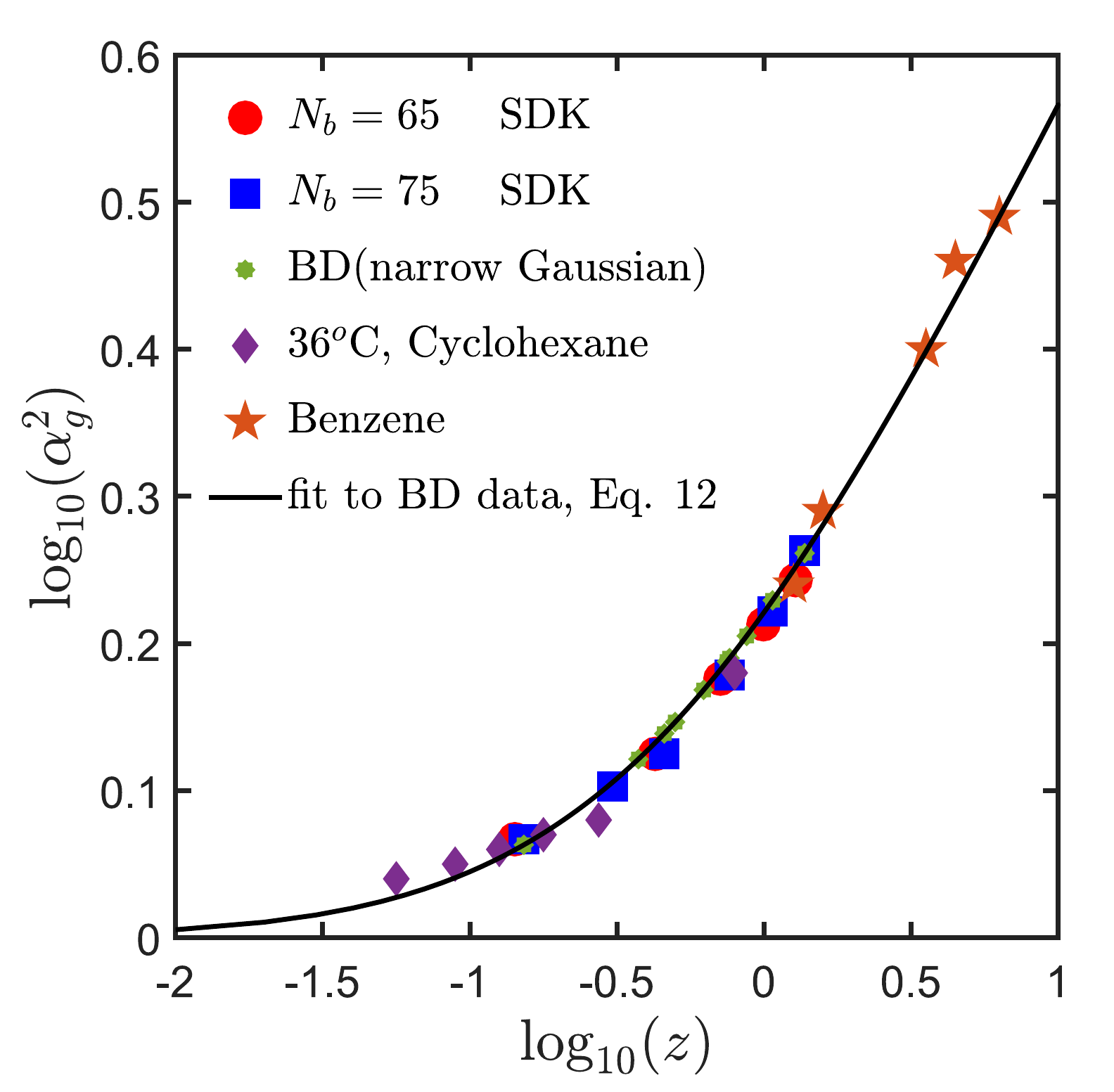}}
	\end{center}
	\caption{(Color online) Universal swelling behaviour of the radius of gyration, $\alpha_g^2$, as a function of the solvent quality, $z$. The red filled circles and the blue filled squares are simulation results with the SDK potential, which is compared with BD simulations obtained with the narrow Gaussian potential\cite{KumarRavi}, and with experimental results for polystyrene in two solvents\cite{Fujita}. The solid line represents the curve fit corresponding to Eq.~(\ref{RGeq}).}
\label{fig:Uniswell}
\end{figure}

\begin{figure}[tbh]
	\begin{center}
	  {\includegraphics[width=3.5in,height=!]{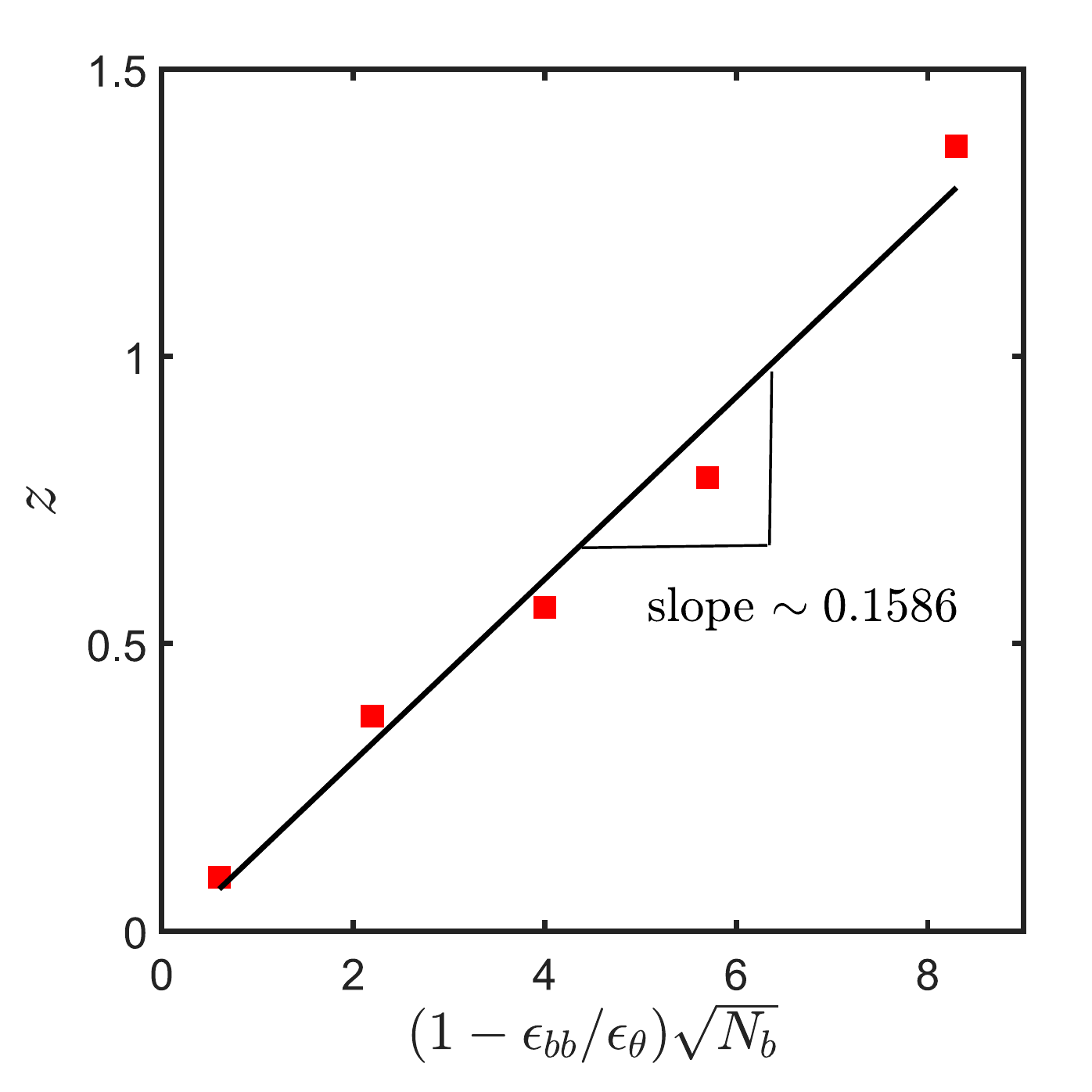}}
	\end{center}
	\caption{Solvent quality $z$ versus the factor $(1-\epsilon_{bb}/\epsilon_{\theta})\sqrt{N_b}$ for polymer chain interacting with SDK potential. The symbols are the simulation data and the straight line gives a linear fit through the data points with slope 0.1586.}
\label{fig:linshift}
\end{figure}

\begin{figure*}[t]
	\begin{center}
		\begin{tabular}{cc}
		    \hspace{-0.8cm}
			\resizebox{9.2cm}{!} {\includegraphics[width=4cm]{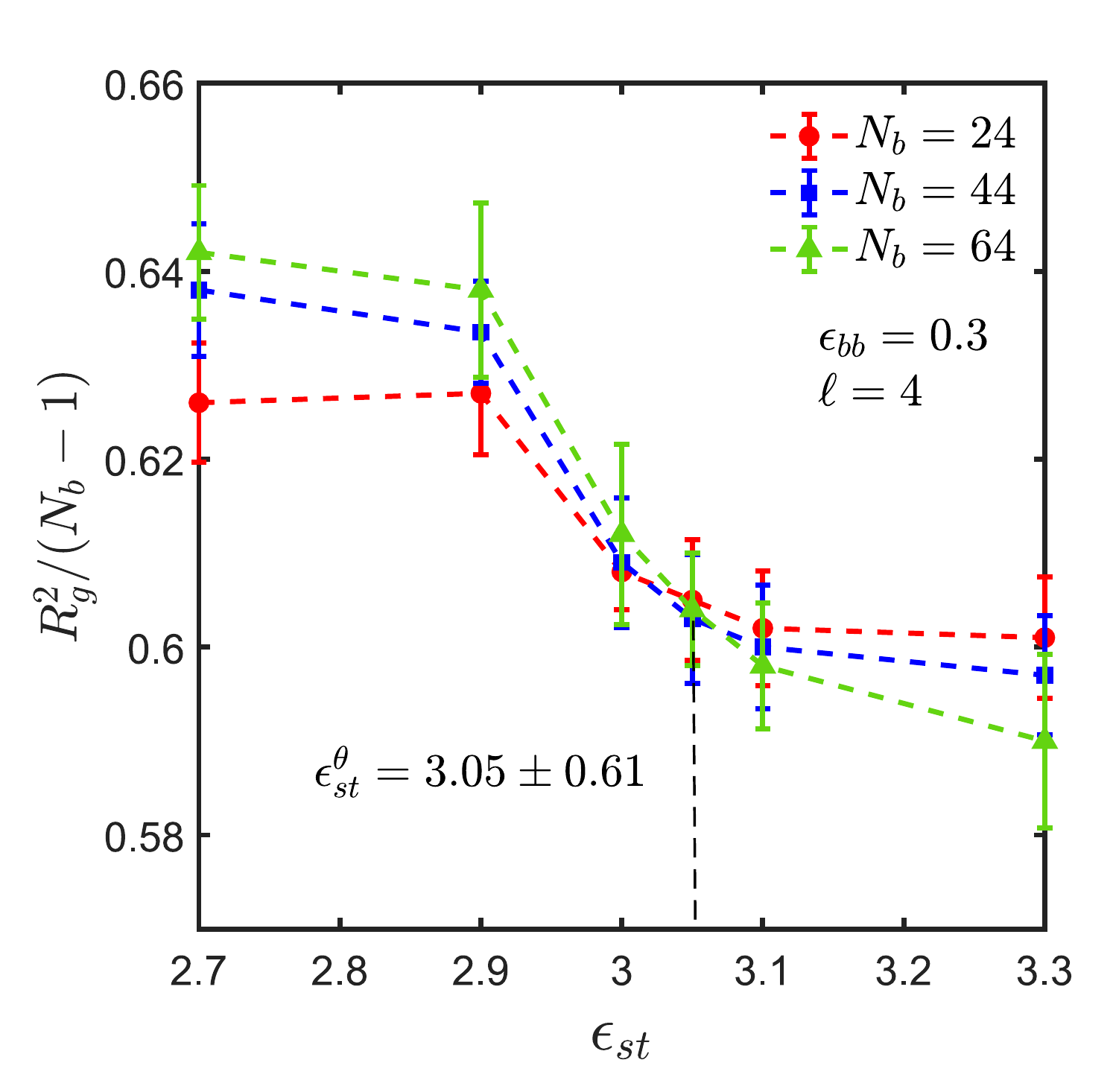}} &
			\hspace{-0.6cm}
			\resizebox{9.2cm}{!} {\includegraphics[width=4cm]{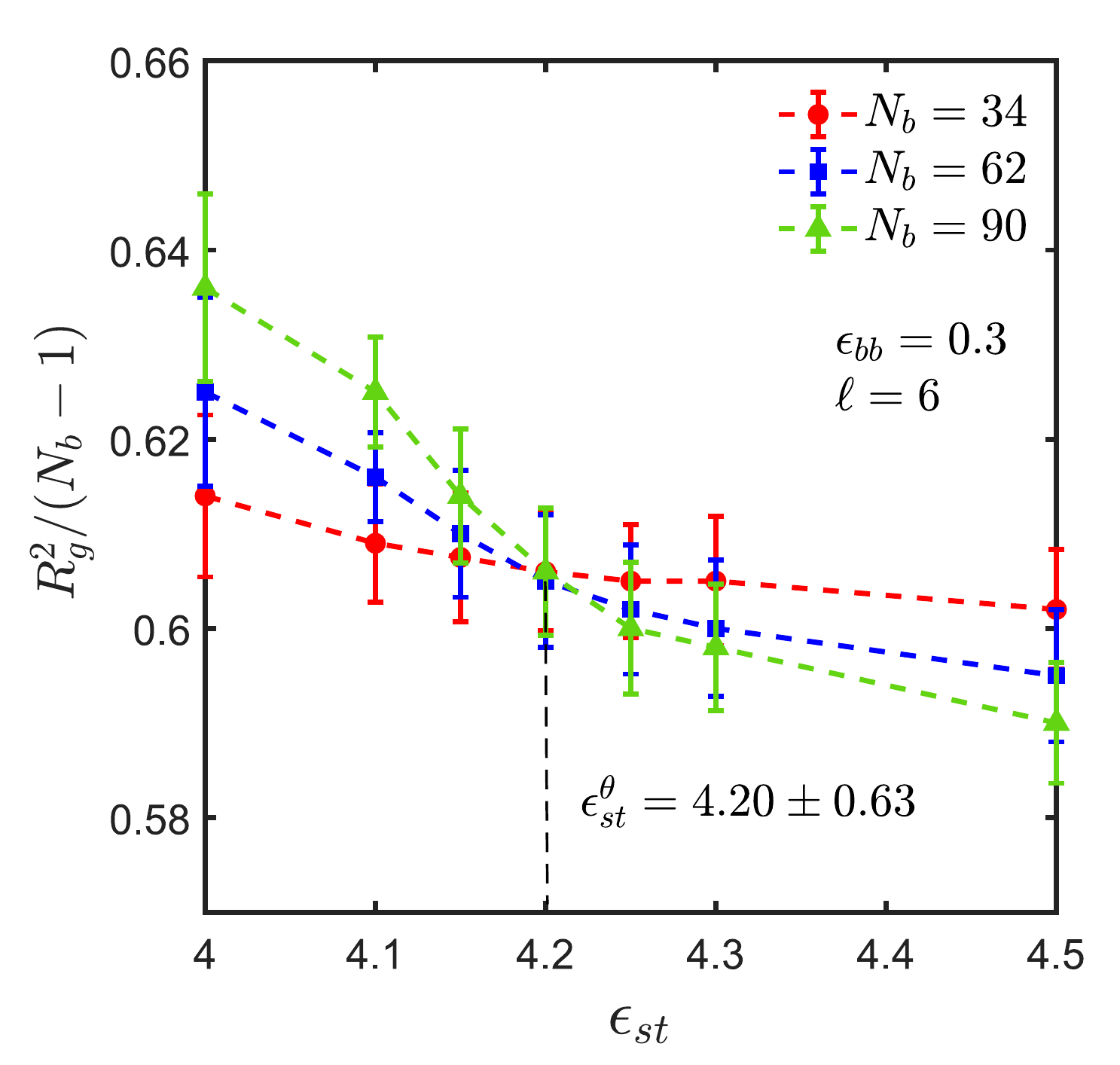}}\\
			(a) & (b)  \\
			\hspace{-0.8cm}
			\resizebox{9.2cm}{!} {\includegraphics[width=4cm]{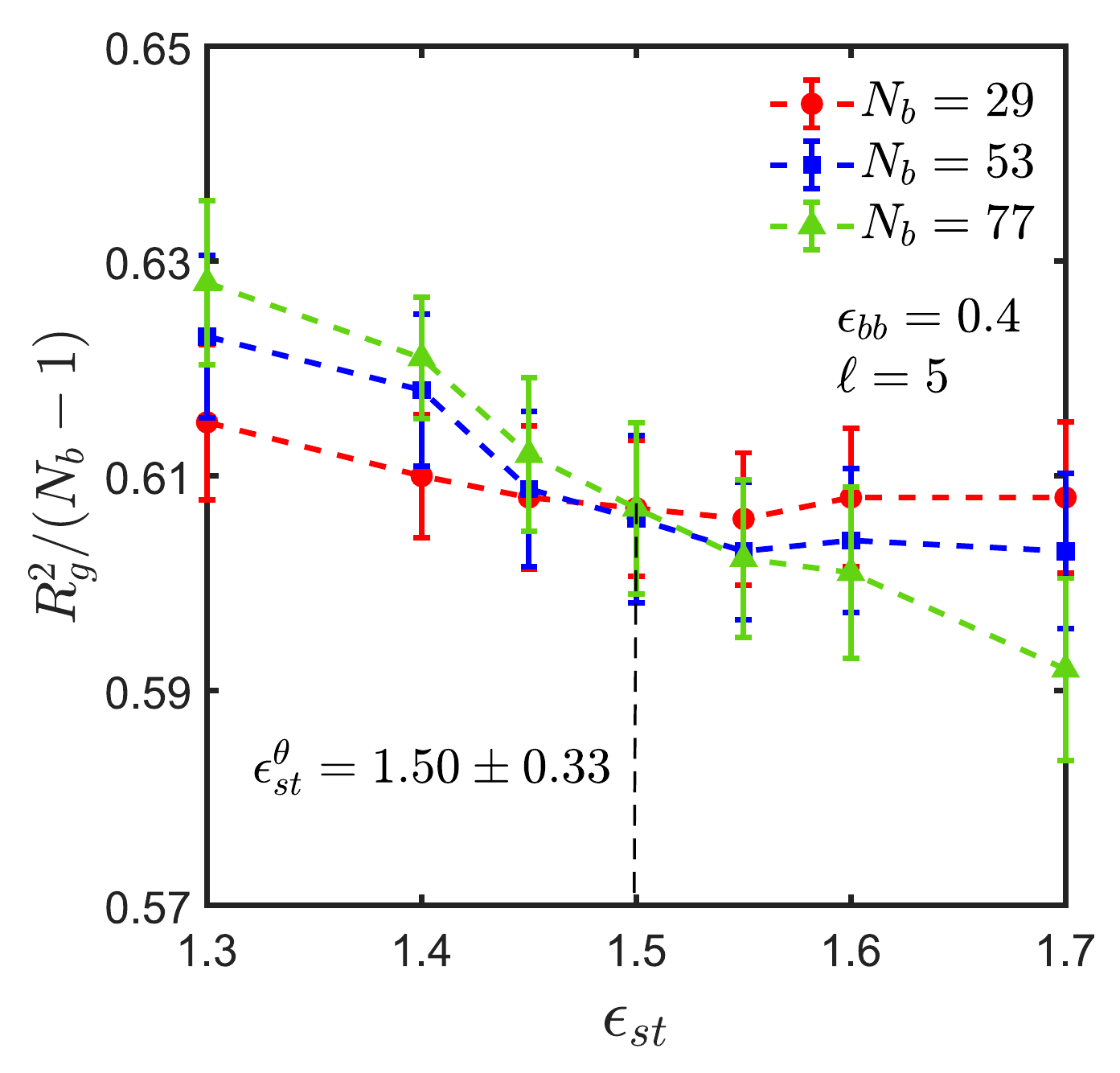}} &
			\hspace{-0.6cm}
			\resizebox{9.2cm}{!} {\includegraphics[width=4cm]{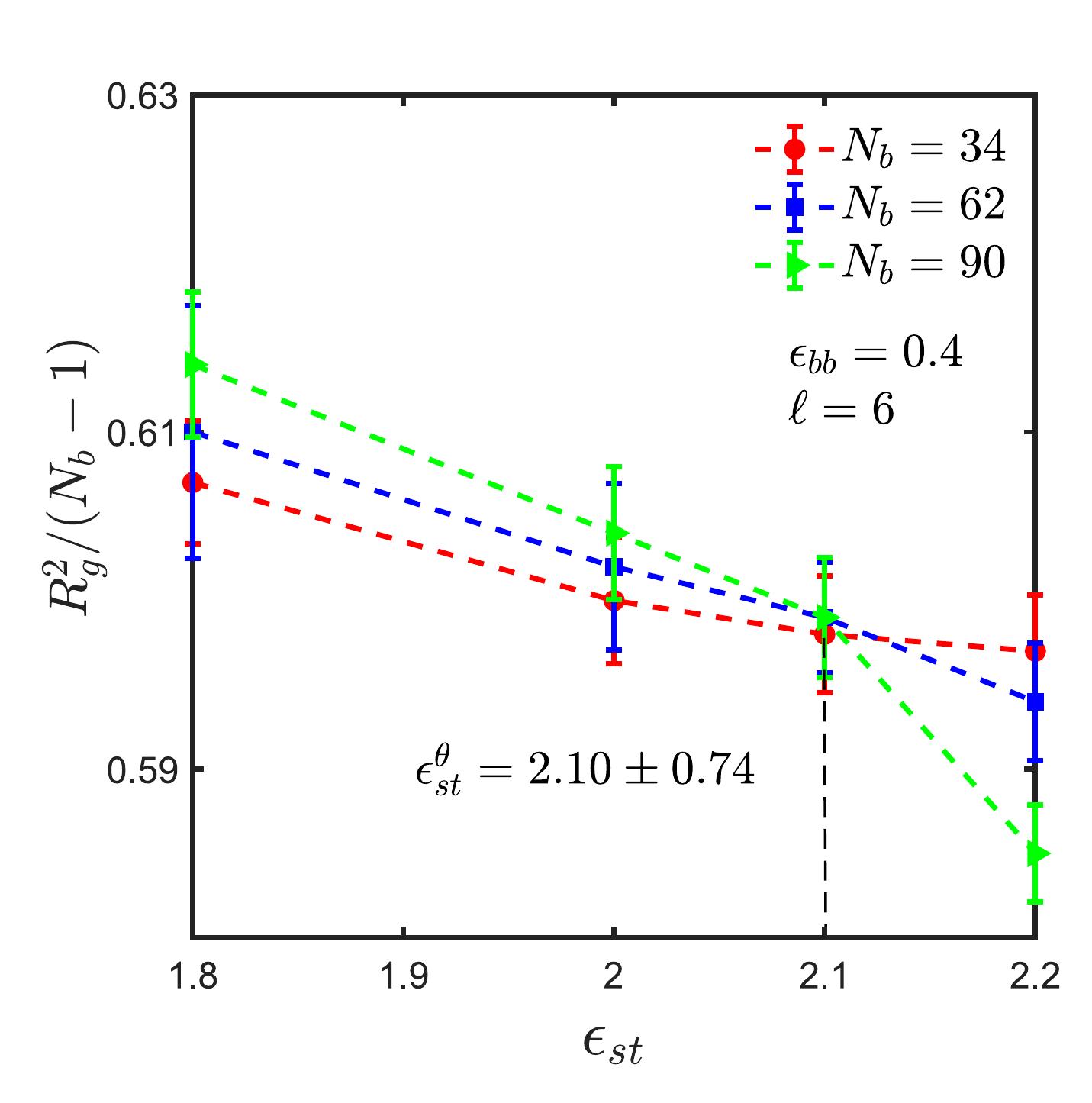}}\\
			(c) & (d)  \\
		\end{tabular}
	\end{center}
	\vskip-15pt
	\caption{ (Color online) The ratio $R_g^2/(N_b-1)$ as a function sticker strength, $\epsilon_{st}$, for a single sticky polymer chain with backbone monomer attraction strengths, $\epsilon_{bb} = 0.3$ and $0.4$, respectively, and spacer length, $\ell=4$, $5$ and $6$, as indicated in the various figure legends. In all the cases the stickers associate with functionality equal to 1. The errorbars for $\epsilon_{st}$ at the point of intersection are estimated by an error propagation scheme discussed in Section~\ref{sec:thetaCalc}.}
    \label{fig:thetpt_SP}
\end{figure*}

In order to discuss the universal swelling of homopolymers, it is necessary to first determine the radius of gyration of a homopolymer under $\theta$-conditions. This in turn requires the determination of the well depth $\epsilon_{bb} = \epsilon_\theta$ of the SDK potential that corresponds to the $\theta$-temperature for homopolymers. This question is taken up in the Appendix, where the value of $\epsilon_\theta$ is estimated by two means, first by determining the value of  $\epsilon_{bb}$ at which the chain obeys random walk statistics, and second by determining the value of $\epsilon_{bb}$ that leads to the second osmotic virial coefficient being zero. As shown in detail in the Appendix, for a SDK potential with cut-off radius $r_c = 1.82 \,\sigma$, we find that the $\theta$-point occurs at $\epsilon_{bb}=\epsilon_\theta = 0.45$. 

The swelling of homopolymers interacting with an SDK potential as the source of the excluded volume force is investigated in this section. The results are compared with the swelling of experimental polymer-solvent systems and earlier predictions of BD simulations, where the excluded volume interactions are modelled by a narrow Gaussian potential given by\cite{Ottinger1996,RaviExclu}
\begin{equation}
 E(\mathbf{r}_{\mu\nu}) = \left(\frac{z^{\ast}}{{d^{\ast}}^3}\right)k_BT\exp\left\lbrace -\frac{1}{2}\frac{\mathbf{r}_{\mu\nu}^2}{{d^{\ast}}^2}\right\rbrace
\end{equation}
\noindent Here, $\mathbf{r}_{\mu\nu}=(\mathbf{r}_{\mu}-\mathbf{r}_{\nu})$, $d^{\ast}$ is a non-dimensional parameter that measures the range of interactions, and $z^{\ast}$ is the non-dimensional strength of excluded volume interactions. In the context of the narrow Gaussian potential, the solvent quality is defined by $z=z^{\ast}\sqrt{N_b}$, which takes into account the dependence on both temperature and chain length. Kumar and Prakash performed BD simulations with the narrow Gaussian potential to obtain the universal swelling ratio as a function of solvent quality $z$\cite{KumarRavi}. Basically, they obtained $\alpha_g^2$ at a particular value of $z$ by carrying out simulations for different chain lengths $N_b$, where the parameter $z^{\ast}$  was evaluated  using the expression $z^{\ast}=z/\sqrt{N_b}$, for each choice of $N_b$. The data accumulated for various values of $N_b$ was then extrapolated to the limit of $N_b\rightarrow\infty$ to obtain the asymptotic $\alpha_g^2$ value, at the chosen value of $z$. The results are plotted in Fig.~\ref{fig:Uniswell} and fitted with an expression suggested earlier by renormalisation group calculations~\cite{Schafer,Freed,desCloz}
\begin{align}\label{RGeq}
 \alpha_g^2 = (1+az+bz^2+cz^3)^m
\end{align}
with fit parameters $a=9.528$, $b=19.48\pm 1.28$, $c=14.92\pm 0.93$ and $m=0.133913\pm 0.0006$\cite{Yamakawa1971,Schafer,KumarRavi}. The fitted curve is the universal thermal crossover swelling curve predicted by BD, and acts as a reference for collapsing swelling data for a range of polymer-solvent systems as discussed below.

Kumar and Prakash~\cite{KumarRavi} showed that experimental data acquired previously~\cite{Fujita} for $\alpha_g^2$ versus $z$, in a variety of different polymer-solvent systems, could also be described by the same universal curve. This is done by defining the experimental solvent quality by $z=k_{\text{expt}}\,\hat{\tau}\sqrt{M}$, where $\hat{\tau}=1-(T_{\theta}/T)$ and adjusting $k_{\text{expt}}$, which is a chemistry dependent constant, in order to achieve data collapse~\cite{KumarRavi}. Swelling data~\cite{Fujita} for polystyrene in cyclohexane at $36^\circ$C, and in benzene at $25^\circ$C and $30^\circ$C, obtained in this manner are shown in Fig.~\ref{fig:Uniswell}.   
We have adopted a similar approach to check whether polymer chains with the SDK potential follow the same universal swelling behaviour. The solvent quality is defined here in terms of the potential well depth as 
\begin{align}\label{Eq:zHP}
 z = k_{\text{SDK}} \left(1-\frac{\epsilon_{bb}}{\epsilon_{\theta}}\right)\sqrt{N_b}
\end{align}     
where $k_{\text{SDK}}$ is a constant dependent on the interaction potential and $\hat{\tau}_{\text{SDK}}=(1-\epsilon_{bb}/\epsilon_{\theta})$ is equivalent to the temperature dependent term, $\hat{\tau}$, defined earlier. Note that the factor $(1-\epsilon_{bb}/\epsilon_{\theta})$ is defined in such a way that in the limit of a $\theta$-solvent its value is zero, while in the good solvent limit ($\epsilon_{bb}=0$), $\hat{\tau}_{\text{SDK}} = 1$. The value of $k_{\text{SDK}}$ is obtained by the following procedure. Simulations are carried out for different values of chain length $N_b$ and well depths $\epsilon_{bb}$, and the mean-squared radius of gyration $R_g^2$ is calculated in each case. The swelling, $\alpha^2_g$, relative to the size of the chain under $\theta$-solvent conditions,  $R_{g\theta}^2$ (obtained from a simulation with $\epsilon_{bb}=\epsilon_{\theta}=0.45$), is calculated in each case, and the corresponding values of $z$ are determined from the universal swelling curve given by Eq.~(\ref{RGeq}). Finally, values of $z$ obtained in this manner are plotted as a function of $(1-\epsilon_{bb}/\epsilon_{\theta})\sqrt{N_b}$, and the resultant curve is fitted with a straight line as shown in Fig.~\ref{fig:linshift}. From the slope one finds $k_{\text{SDK}} = 0.1586$.

The swelling of polymer chains with $N_b=65$ and $N_b=75$, obtained with the SDK potential for a set of values of $z$ obtained in this manner, is compared in Fig.~\ref{fig:Uniswell} with earlier results from BD simulations and experimental measurements of the swelling of polystyrene in cyclohexane and benzene. It is clear that the SDK potential reproduces the universal swelling behaviour in the thermal crossover regime obtained previously with the narrow Gaussian potential. It is worth noting here, however, that in the case of the SDK potential, we have not extrapolated finite chain data to the long chain limit, as was done for the narrow Gaussian potential. We found that this was unnecessary since the results for $N_b=65$ and $N_b=75$ were already lying on the universal curve.

\begin{figure}[tbh]
	\begin{center}
	 {\includegraphics[width=3.5in,height=!]{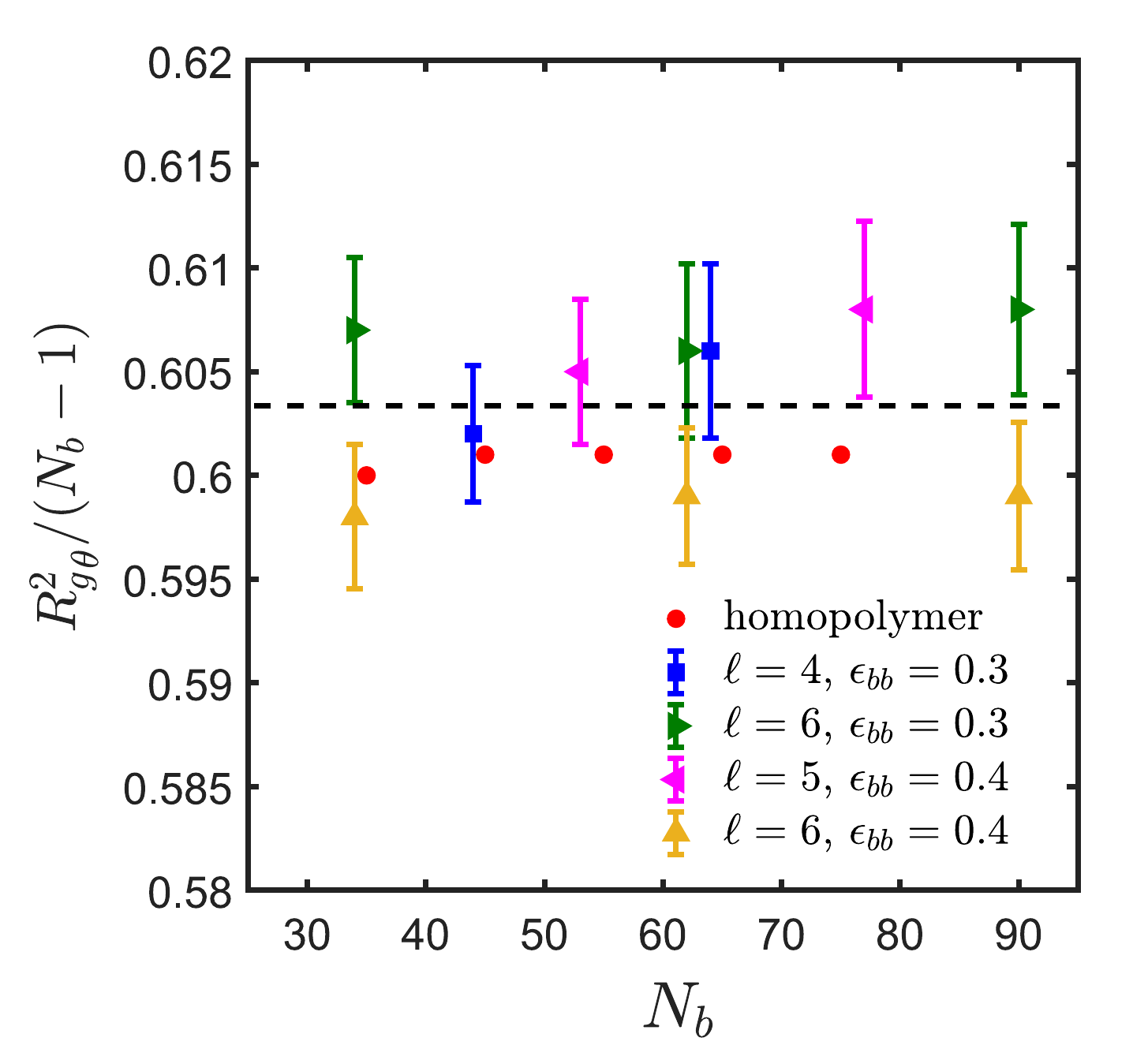}}
	\end{center}
	\caption{(Color online) The ratio $R_g^2/(N_b-1)$ at the $\theta$-point, denoted by $R_{g_{\theta}}^2/(N_b-1)$, for different chain lengths, $N_b$, spacer monomer, $\ell$, and backbone monomer attraction strength, $\epsilon_{bb}$. The dashed line corresponds to the constant value of the ratio $R_{g_{\theta}}^2/(N_b-1)$, which is estimated to be 0.603. }
\label{fig:Rg2theta}
\end{figure}  

\section{\label{sec:SwellSP}$\theta$-point and swelling of sticky polymers}

As mentioned earlier, the inclusion of sticky groups decreases the effective solvent quality. The purpose of this section is to present results which establish, within the studied parameter range, that the collapse of the sticky polymer chain is a standard second order $\theta$ transition. We show this by demonstrating that all the methods that have been applied to homopolymers, both for the localisation of the $\theta$ transition (as discussed in the Appendix), and for the study of the universal scaling for the swelling (as discussed in the previous section) can be carried over. 

\subsection{\label{sec:thetaCalc}$\theta$-point for sticky polymer solutions}
\subsubsection{\label{sec:Rg2scaling} Scaling of the radius of gyration}
For the sticky chain, we keep $\epsilon_{bb}$ and $\ell$ fixed, and study $R_g^2$ as a function of $\epsilon_{st}$, which we use as the control parameter that drives the transition. Following the procedure described in Appendix A for homopolymers, intersection plots for $R_g^2/(N_b-1)$ versus $\epsilon_{st}$ are presented in Fig.~\ref{fig:thetpt_SP}, which allows us to find the $\theta$-point for various choices of $\epsilon_{bb}$ and $\ell$, $\epsilon_{st}^{\theta} = \epsilon_{st}^{\theta}(\epsilon_{bb}, \ell)$. The error in $\epsilon_{st}^{\theta}$ is estimated by linear interpolation between the data points adjacent to the intersection, combined with standard error propagation. We will show shortly that for our model, $\epsilon_{st}^{\theta}$ can be determined by an alternative simpler and perhaps more accurate procedure.

\begin{table}
\begin{center}
\setlength{\tabcolsep}{10pt}
\renewcommand{\arraystretch}{1.5}
\begin{tabular}{| c | c | c | c |}
\hline
 $\ell$     & $\epsilon_{bb}$ 
            & $\epsilon_{st}^{\theta}$ (from $R_g^2$ scaling)
            & $\epsilon_{st}^{\theta}$ (from $B_2$)   
\\                
\hline
\hline
            $4$
            & $0.3$
            & $3.05 \pm 0.61$
            & $3.231 \pm 0.012$  
\\
            $5$
            & $0.4$
            & $1.50 \pm 0.33$
            & $1.215 \pm 0.065$ 
\\
            $6$
            & $0.3$
            & $4.20 \pm 0.63$
            & $-$
\\
            $6$
            & $0.4$
            & $2.10 \pm 0.74$
            & $-$
\\
\hline
\end{tabular}
\end{center}
\caption{Comparison between the $\theta$-points estimated from the scaling of radius of gyration and second virial coefficient for solutions of sticky polymers with different spacer lengths, $\ell$, and backbone monomer interaction strengths, $\epsilon_{bb}$.}
\label{tab:thetapt_table}
\end{table}

It is interesting to note that the ratio $R_g^2/(N_b-1)$ at the $\theta$-point, denoted by $R_{g_{\theta}}^2/(N_b-1)$, assumes a constant value, as shown in Fig.~\ref{fig:Rg2theta}, for monomers interacting via the SDK potential, irrespective of the spacer length and backbone solvent quality. It is also evident that the value of the ratio $R_{g_{\theta}}^2/(N_b-1)$ is the same for both homopolymers and sticky polymers, suggesting that the Kuhn step length is the same in both cases, and is independent of the presence of sticky groups. This implies that one does not need to do simulations for each and every system in order to estimate the $\theta$-point and calculate the corresponding $R_{g_{\theta}}^2$. This, of course, simplifies matters substantially. It is worth emphasising, however, that this is most probably a special feature of our model, and probably will no longer be true for chemically more realistic models.

\subsubsection{\label{sec:thetapt2}Calculation of the second virial coefficient}  
For sticky polymers, the estimation of the $\theta$-point  from the second virial coefficient was carried out as was done for homopolymers (see Appendix B), with a set of about $5 \times 10^7 - 10^8$ configurations of chain pairs. Since the Boolean variables $q_{\mu\nu}$ (see Section~\ref{sec:SM}) were not stored in the course of the simulations, it was necessary in situations where there were three stickers within interaction distance of each other, to define $q_{\mu\nu}$ by some rule. For simplicity, we picked the two stickers with the lowest monomer indices to be bonded. For the rare case of four or more stickers, we proceeded in an analogous fashion. 

\begin{figure*}[t]
	\begin{center}
		\begin{tabular}{cc}
		    \hspace{-0.8cm}
			\resizebox{9.2cm}{!} {\includegraphics[width=4cm]{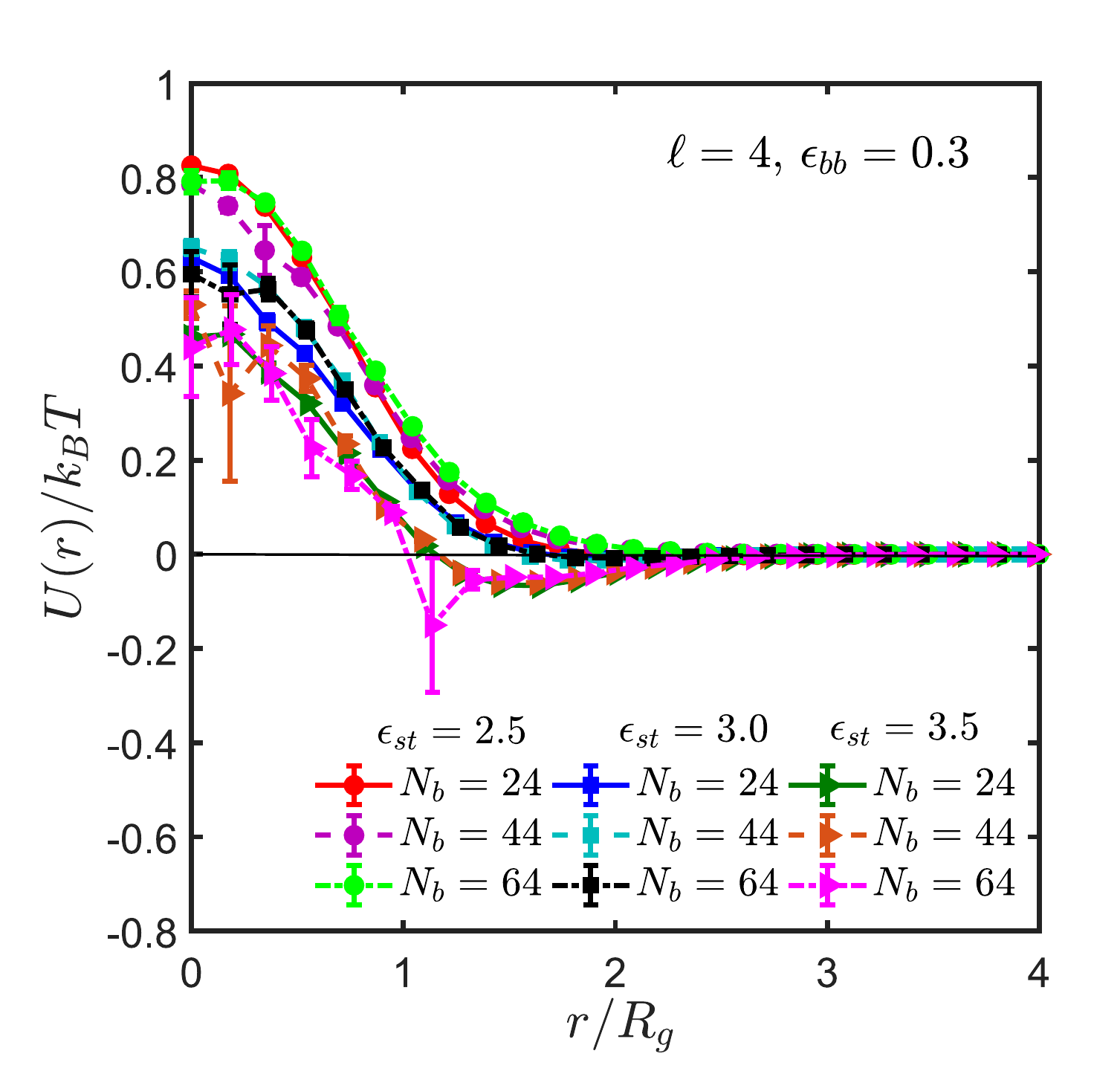}} &
			\hspace{-0.6cm}
			\resizebox{9.2cm}{!} {\includegraphics[width=4cm]{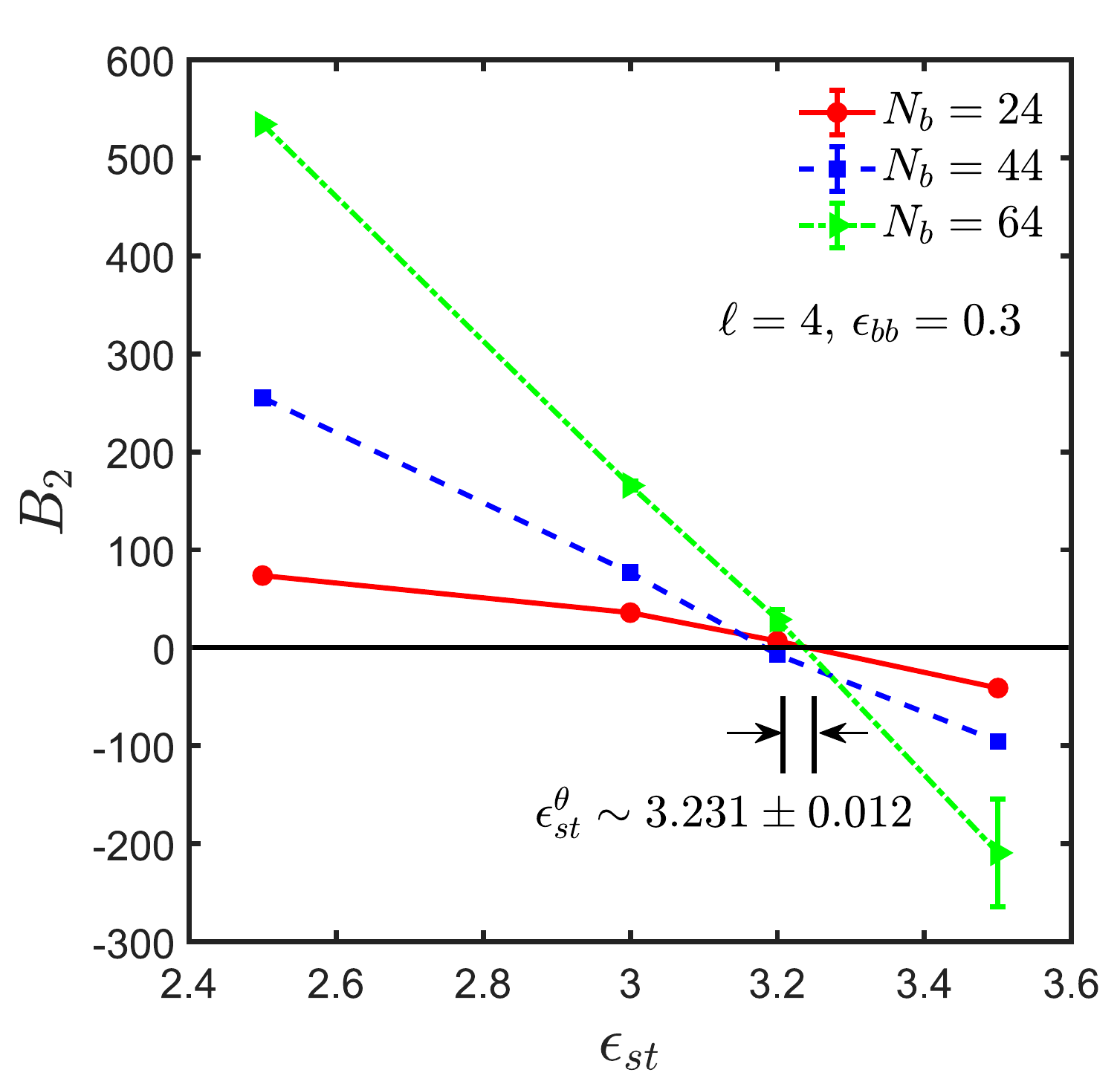}}\\
			(a) & (b)  \\
			\hspace{-0.8cm}
			\resizebox{9.2cm}{!} {\includegraphics[width=4cm]{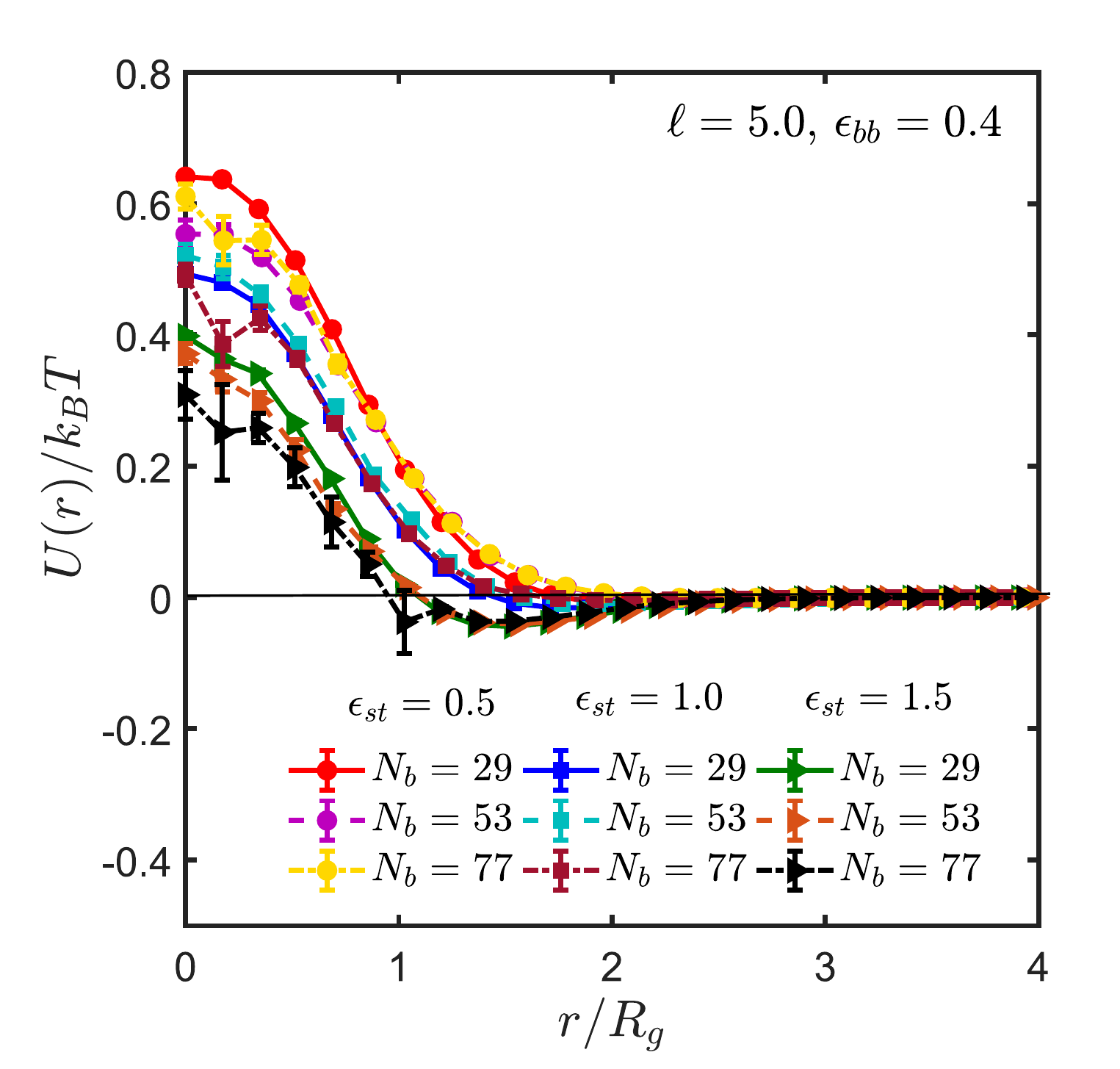}} &
			\hspace{-0.6cm}
			\resizebox{9.2cm}{!} {\includegraphics[width=4cm]{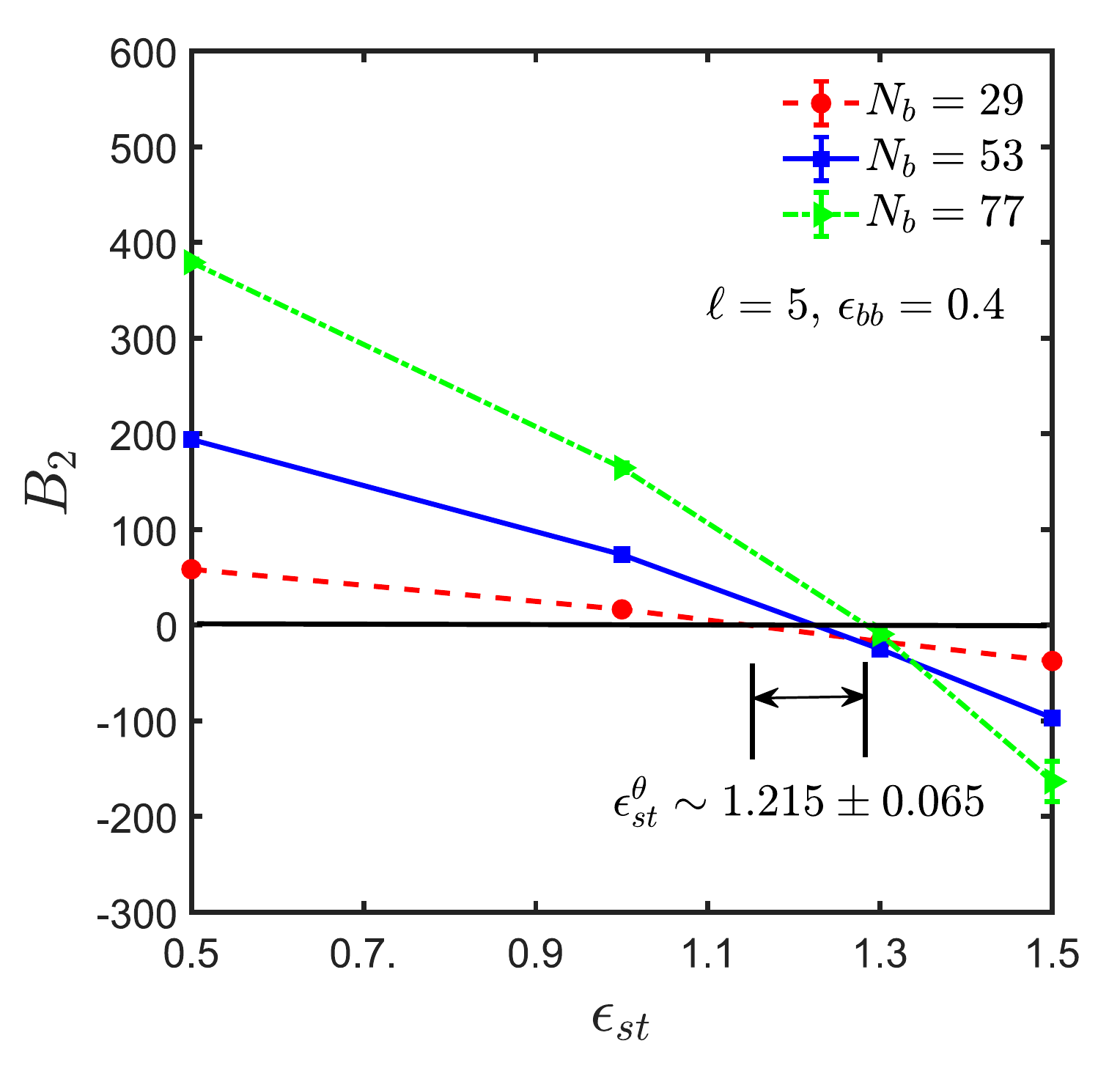}}\\
			(c) & (d)  \\
		\end{tabular}
	\end{center}
	\vskip-15pt
	\caption{(Color online) ((a) and (c)) The effective potential, $U(r)/k_BT$, as a function of the separation distance, $r/R_g$, between the centres of mass of a pair of sticky polymer chains at different values of sticker strength, $\epsilon_{st}$. ((b) and (d)) The second osmotic virial coefficient, $B_2$, as a function of the sticker strength, $\epsilon_{st}$, for different chain lengths. The estimated value of the $\theta$-point, $\epsilon_{st}^{\theta}$, from the second virial coefficient is  $3.231 \pm 0.012$ and $1.215 \pm 0.065$ for $\ell=4$, $\epsilon_{bb}=0.3$ and $\ell=5$, $\epsilon_{bb}=0.4$, respectively.}
    \label{fig:StkpolyB2}
\end{figure*}

Figures~\ref{fig:StkpolyB2} shows the effective interaction potentials and second virial coefficients for systems with spacer lengths $\ell = 4, 5$, and $\epsilon_{bb} = 0.3, 0.4$. One sees that even though the method is hampered by similar problems as in the homopolymer case, it is nevertheless possible to locate the $\theta$-point with reasonable accuracy, which is actually significantly better than that obtained from the scaling of $R_g^2$. The results are summarised in Table \ref{tab:thetapt_table}. With increase in spacer length ($\ell=6$) and sticker strength, the sampling gets poorer and less efficient, as explained in Appendix B, such that it was not possible to determine the $\theta$-point by the virial coefficient method satisfactorily. 

\subsubsection{\label{sec:thetasurf}The $\theta$-surface for sticky polymers}

\begin{figure}[tbh]
	\begin{center}
	    {\includegraphics[width=3.5in,height=!]{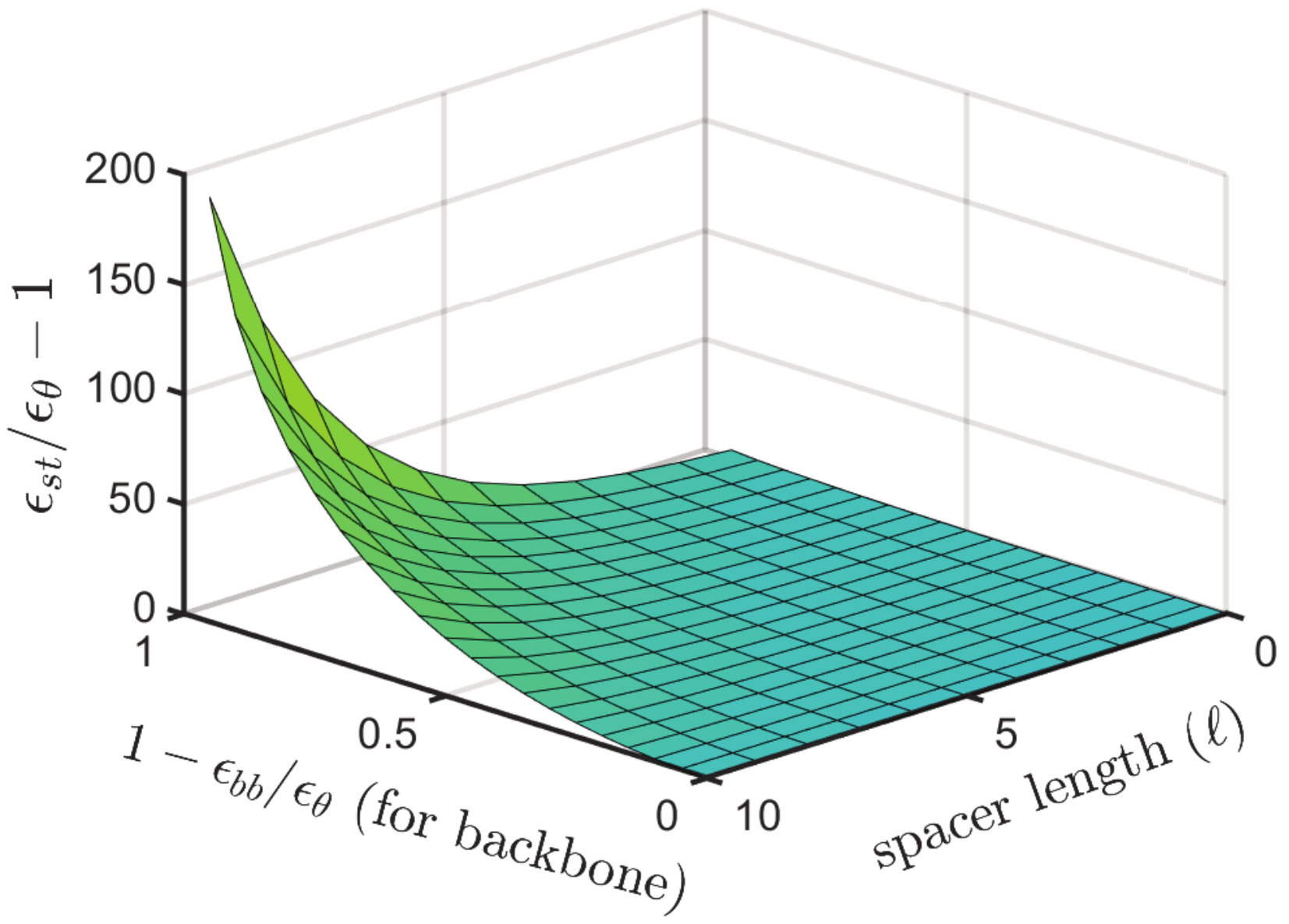}}
	\end{center}
	\caption{(Color online) Schematic of a $\theta$-surface in terms of the scaled variables $(\epsilon_{st}/\epsilon_{\theta}-1)$ plotted against the scaled backbone solvent quality, ($1-\epsilon_{bb}/\epsilon_{\theta}$), and spacer length, $\ell$. Points above the surface indicate solutions of sticky polymers in poor solvent whereas points below the surface represents solutions in a good solvent.}
\label{fig:thetasurface}
\end{figure}

Since the location of the $\theta$-point, $\epsilon_{st}^{\theta}$, depends on both $\ell$ and $\epsilon_{bb}$, the full phase diagram of the system is three-dimensional, with a two-dimensional surface of $\theta$-points separating the good and poor solvent regions. Figure~\ref{fig:thetasurface} is a schematic representation of such a surface, where we confine attention to the physically interesting case $\epsilon_{st} > \epsilon_{bb}$. This implies that under $\theta$ conditions for the chain as a whole, the backbone is in a relatively good solvent ($\epsilon_{bb}<\epsilon_{\theta}$), meaning that the conformations are the result of a competition between backbone-backbone repulsion and sticker-sticker attraction. 

\subsection{\label{sec:UniswellSP}Universal swelling of sticky polymers}

In order to study the universal swelling for the sticky chain, a suitable definition of the solvent quality parameter $z$ is required. Based upon the observation that we have chosen $\epsilon_{st}$ as the independent control parameter that drives the transition, it is logical to generalise Eq.~(\ref{Eq:zHP}) as  
\begin{align}\label{Eq:zSP}
 z = g(\ell,\epsilon_{bb})\left(1-\frac{\epsilon_{st}}{\epsilon_{st}^{\theta}(\ell,\epsilon_{bb})}\right)\sqrt{N_b}
\end{align}      
where $g(\ell,\epsilon_{bb})$ is a material dependent function of spacer length ($\ell$) and backbone monomer interaction strength ($\epsilon_{bb}$). In the limit of $\epsilon_{st}=\epsilon_{bb}$, the effective solvent quality becomes the same as that of the corresponding homopolymer, and the sticky chain becomes indistinguishable from it (see Eq.(\ref{Boolean})). By definition, for $\epsilon_{st}=\epsilon_{st}^{\theta}$ the overall solvent quality $z=0$.

\begin{figure}[tbh]
	\begin{center}
	  {\includegraphics[width=3.5in,height=!]{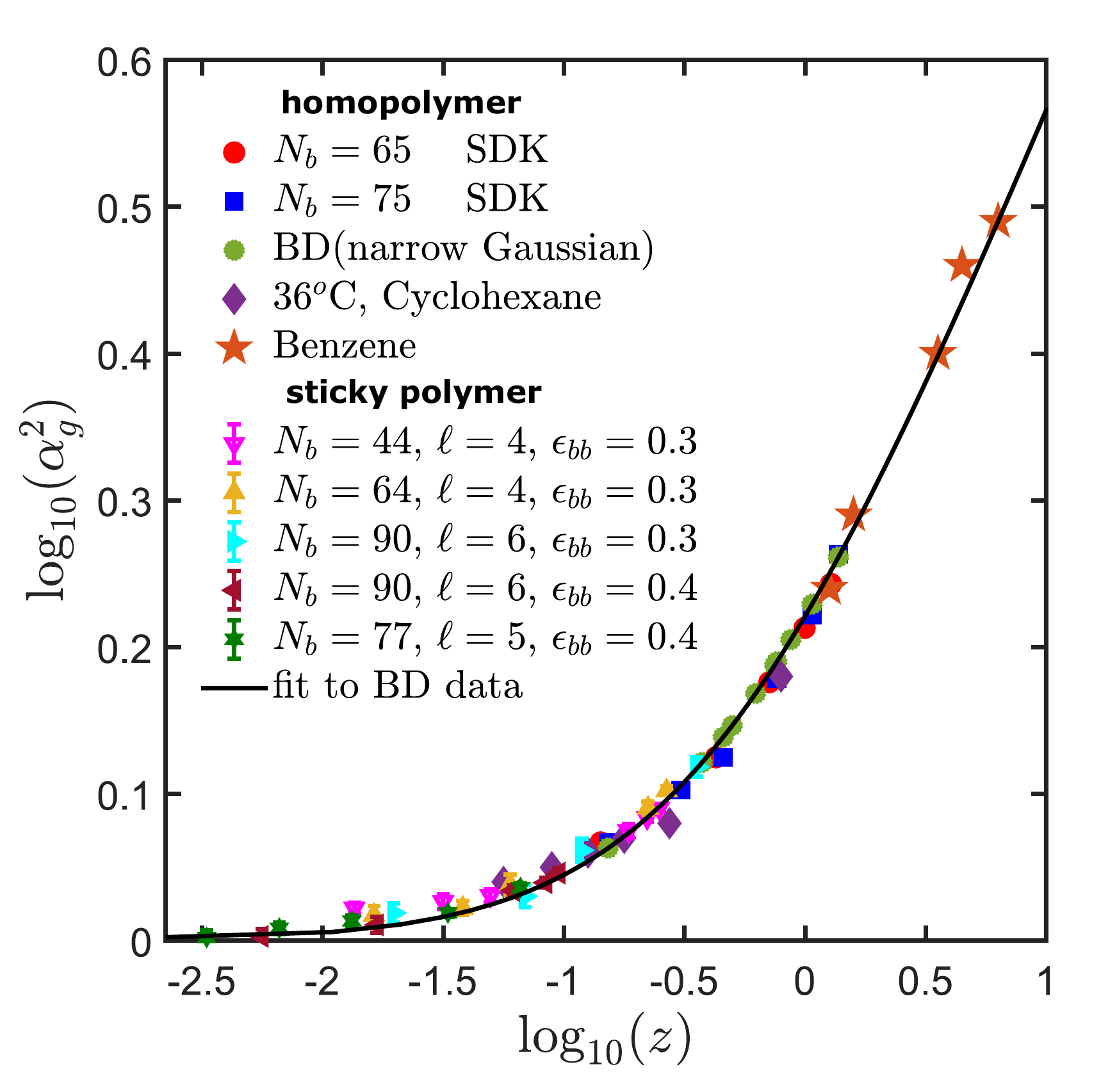}}
	\end{center}
	\caption{(Color online) Universal swelling of the radius of gyration, $\alpha_g^2$, as a function of the solvent quality, $z$. The swelling of the sticky polymers with different spacer lengths, $\ell$, and backbone monomer attraction strengths, $\epsilon_{bb}$, is compared with the swelling of homopolymers, as indicated in the figure. The sticker functionality is equal to 1 in all cases. The solid line represents the curve fit corresponding to Eq.~(\ref{RGeq}).}
\label{fig:Uniswell_SP}
\end{figure}

For obtaining the swelling curve for sticky polymers, one can exploit the simplification that comes from the  constancy of the ratio $R_{g_{\theta}}^2/(N_b-1)$, independent of all other parameters, which renders the need for extra simulations to calculate $R_{g_{\theta}}^2$ superfluous. Simulations are carried out for different values of $\ell$, $\epsilon_{bb}$,  $\epsilon_{st}$, and $N_b$, and the swelling ratio, $\alpha_g^2$, is then calculated for each set of these parameters. For values of $\ell$ and $\epsilon_{bb}$ for which $\epsilon_{st}^{\theta}$ is known, $g(\ell,\epsilon_{bb})$ can be estimated using the same technique as was used to find $k_{\text{SDK}}$ in the case of homopolymers, discussed in Section~\ref{sec:UniSwellHP}. The results are tabulated in Table~\ref{tab:k_stk_table}. Plotting $\alpha_g^2$ versus $z$, as displayed in Fig.~\ref{fig:Uniswell_SP}, shows that the swelling of sticky polymers relative to its $\theta$ state follows the same universal curve already presented in Fig.~\ref{fig:Uniswell} for homopolymer systems.

\begin{table}
\begin{center}
\setlength{\tabcolsep}{10pt}
\renewcommand{\arraystretch}{1.5}
\begin{tabular}{| c | c | c |}
\hline
 $\ell$     & $\epsilon_{bb}$ 
            & $g(\ell,\epsilon_{bb})$  
\\                
\hline
\hline
            $4$
            & $0.3$
            & $0.0415 \pm 0.0021$
\\
            $6$
            & $0.3$
            & $0.0437 \pm 0.0024$
\\
            $5$
            & $0.4$
            & $0.0113 \pm 0.0011$
\\
            $6$
            & $0.4$
            & $0.0124 \pm 0.0011$
\\
\hline
\end{tabular}
\end{center}
\caption{Estimated values of the function $g(\ell,\epsilon_{bb})$ for different values of spacer length, $\ell$, and backbone monomer interaction strength, $\epsilon_{bb}$.}
\label{tab:k_stk_table}
\end{table}

The existence of the universality of the swelling of sticky polymers, which has been established above, can be used to determine the $\theta$-point $\epsilon_{st}^{\theta}(\ell,\epsilon_{bb})$, along with $g(\ell,\epsilon_{bb})$, for any pair of values of $\ell$ and $\epsilon_{bb}$, without the complicated analysis of subsection~\ref{sec:thetaCalc}. One only needs data for two different $\epsilon_{st}$ values, while $N_b$, $\ell$, and $\epsilon_{bb}$ are being kept fixed. The swelling curve allows us to transform the $\alpha_g^2$ values to corresponding solvent qualities $z$. Therefore, inserting all known parameters into Eq.~(\ref{Eq:zSP}) gives rise to two equations with two unknowns which, after solution, provide the desired values $g(\ell,\epsilon_{bb})$ and $\epsilon_{st}^{\theta}(\ell,\epsilon_{bb})$. A sample calculation to demonstrate the above method is as follows. The parameters chosen are $N_b=64$, $\ell=4$, $\epsilon_{bb}=0.3$ and the two sticker strengths are $\epsilon_{st} = 1.0$ and $2.5$. Considering $R_{g_{\theta}}^2/(N_b-1)=0.603$, the values of $\alpha_g^2$  obtained from the simulations for the given set of parameters are equal to $1.245$ and $1.105$ for $\epsilon_{st}=1.0$ and $\epsilon_{st}=2.5$, respectively. The corresponding values of $z$ are estimated from Eq.~(\ref{RGeq}) to be $0.2635$ and $0.096$ for $\epsilon_{st}=1.0$ and $\epsilon_{st}=2.5$, respectively. Substituting the values of $z$ in Eq.~(\ref{Eq:zSP}) and simultaneously solving the two linear equations for the unknowns, gives $\epsilon_{st}^{\theta}(\ell,\epsilon_{bb})=3.36$ and $g(\ell,\epsilon_{bb}) = 0.047$, which is, within error bars, consistent with the previously estimated values of $\epsilon_{st} (\ell,\epsilon_{bb})= 3.05 \pm 0.61$ and $g(\ell,\epsilon_{bb})=0.0415 \pm 0.0021$ (see Fig.~\ref{fig:thetpt_SP} (a) and Table~\ref{tab:k_stk_table}).
 
\begin{figure}[tbh]
	\begin{center}
	    {\includegraphics[width=3.5in,height=!]{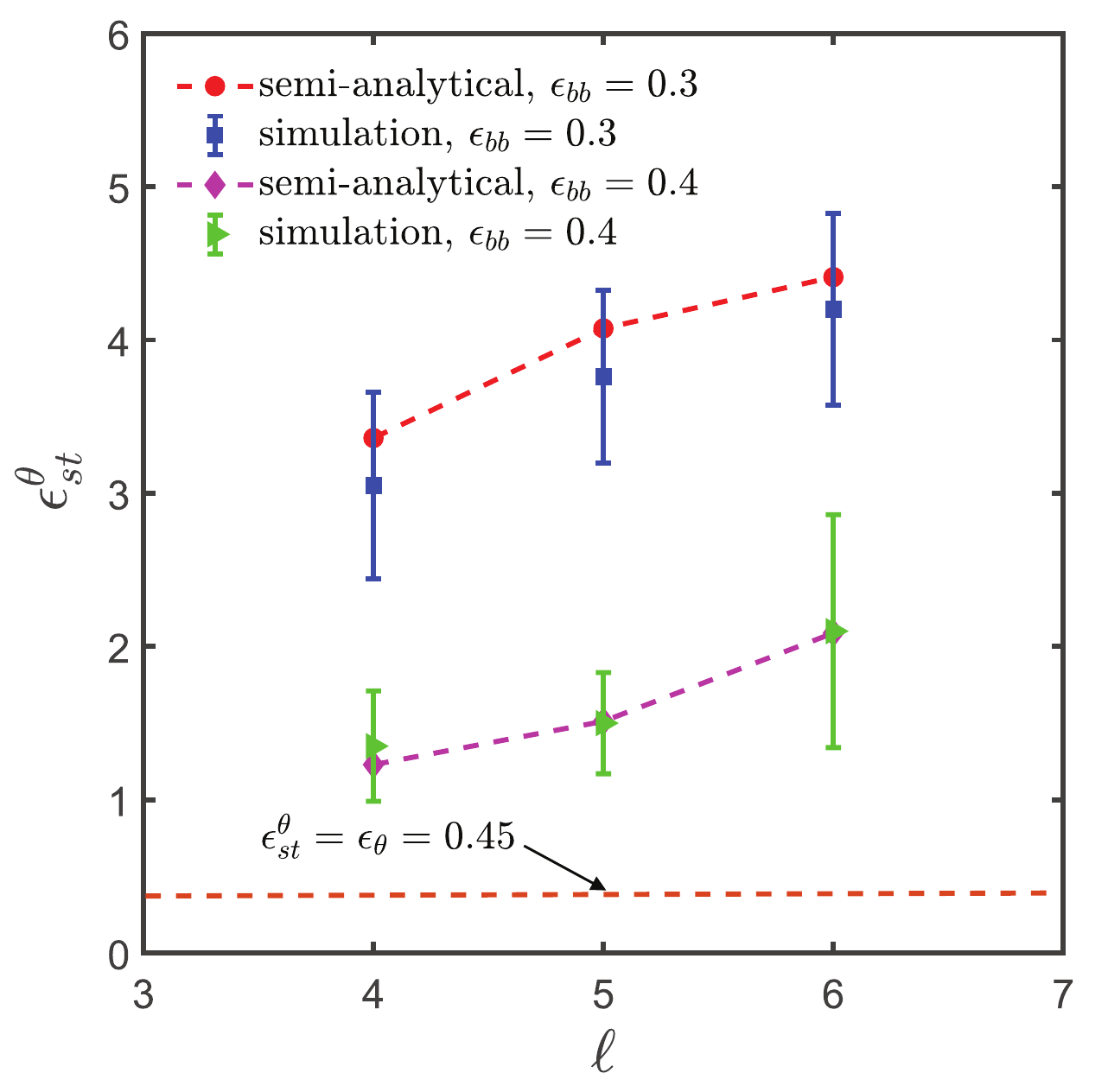}}
	\end{center}
	\caption{(Color online) Sticker strength under $\theta$-solvent conditions, $\epsilon_{st}^{\theta}$, versus the spacer length, $\ell$, for different backbone attraction strengths, $\epsilon_{bb}$. The open symbols represent the elaborate procedure for obtaining $\epsilon_{st}^{\theta}$ described in Section~\ref{sec:thetaCalc}, while the filled symbols are obtained by the semi-analytical procedure described in Section~\ref{sec:UniswellSP}. The dashed line is drawn to guide the eye. The constant straight line indicates the limiting value of $\epsilon_{st}^{\theta}$, as discussed in the context of Eq.~(\ref{Eq:zSP}). }
\label{fig:thpt}
\end{figure}

A comparison of the values of $\epsilon_{st}^{\theta}$ obtained from the semi-analytical estimation procedure discussed above with that from full blown simulations as carried out in the context of Fig.~\ref{fig:thetpt_SP}, for different spacer monomers $\ell$ and backbone solvent qualities $\epsilon_{bb}$, is shown in Fig.~\ref{fig:thpt}. The plot shows a satisfactory agreement between the two methods and implies that the calculation of $\epsilon_{st}^{\theta}$ using Eq.~(\ref{Eq:zSP}) provides a reasonable estimate of the $\theta$-point for the sticky polymer systems. Fig.~\ref{fig:thpt} also suggests, as expected intuitively, that the sticker strength at the $\theta$-point, $\epsilon_{st}^{\theta}$, increases monotonically with spacer length, $\ell$, for a given value of $\epsilon_{bb}$. With an increase in spacer monomers, the sticker density along the polymer backbone decreases, and it requires a much higher attractive strength for the stickers to make the chain follow RW statistics. In all our simulations, the value of the sticker strength $\epsilon_{st}$ is taken to be greater than $\epsilon_{\theta}$, while $\epsilon_{bb} < \epsilon_{\theta}$. The corresponding increase in solvent quality for the backbone is thus compensated by an increased attraction between the stickers.  At $\epsilon_{bb}=\epsilon_{\theta}$, the backbone is in a $\theta$-solvent condition, and under such circumstances the sticker strength at the $\theta$-point, $\epsilon_{st}^{\theta}$, is equal to $\epsilon_{\theta}$, which is the limiting value of $\epsilon_{st}^{\theta}$, indicated by the constant straight line in Fig.~\ref{fig:thpt} and discussed further  in the context of Eq.~(\ref{Eq:zSP}).

\section{\label{conclusions} Discussion and conclusions}

Using the Soddemann-D\"{u}nweg-Kremer potential to model excluded volume interactions, and defining a renormalised solvent quality for sticky polymer solutions, the swelling of the radius of gyration has been shown to be identical to the universal swelling of homopolymers in the thermal crossover regime. Additionally, the Kuhn segment length under $\theta$ conditions, for our model, is found to be the same for chains with and without stickers. This allows, in combination with the known universal swelling curve, a fairly easy determination of the two-dimensional $\theta$-surface embedded in the three-dimensional $(\ell, \epsilon_{bb}, \epsilon_{st})$ phase diagram.

The collapse transition observed here is a standard second-order transition and as a consequence, all scaling laws and universal properties are faithfully reproduced. In the parameter range that we have studied, a strong first-order transition can be ruled out. In hindsight this is perhaps not too surprising, since the only theoretical possibility for the existence of first-order behaviour arises from a strong coupling of the conformational degrees of freedom to the Boolean degrees of freedom that describe functionality. As seen from Eq.~(\ref{Boolean}), the strength of the coupling is directly proportional to $\epsilon_{st} - \epsilon_{bb}$. In the parameter range where we did the simulations, this difference was never very large. For this reason, the possibility of a first-order transition in the opposite limit $\epsilon_{st} - \epsilon_{bb}   \gg 1$ cannot be ruled out within the framework of the current investigation.

One might then consider a situation where the backbone is under very good solvent conditions, $\epsilon_{bb} =0$, while $\epsilon_{st}$ is so large that nevertheless a collapse would occur. Note that in such a situation the transition would be \emph{entropy}-driven rather than \emph{energy}-driven. This is so because the condition $\varphi =1$ for the bonds would lead to a complete saturation, with no residual attraction being left. The reason for a collapsed conformation would then be merely entropic because such a state allows for many more possibilities to form bonds than a swollen chain, which would only allow association of stickers that are near each other on the chain. Note, however, that such a situation would essentially be impossible to simulate with standard Brownian dynamics, simply because the breaking of a once-formed bond would be extremely rare. It might be possible to investigate such a situation with advanced Monte Carlo algorithms like parallel tempering~\cite{earl_parallel_2005}, but this is beyond the scope of the present investigation.

There are a number of previous studies, in the context of models for both synthetic and biological polymer solutions, where a first-order rather than a second-order transition has been observed. For instance, in the model proposed by Jeppesen and Kremer~\cite{Jeppesen1996} for the phase-behaviour of polyethylenoxide in water, each monomer has a Boolean degree of freedom that enters the interaction energy. Depending on the strength of the coupling parameter, they find a second or first-order transition, with the first order transition being entropy driven for reasons similar to those discussed above. More recently, \citet{Scolari2015} have developed a model for the folding of chromosomes due to self-attraction, and the formation of loops due to bridging proteins. Their model is similar to that used here for sticky polymers, with the bridging interactions (which are distributed uniformly along the backbone of the polymer) playing the role of stickers. However, their model differs from the one used here in certain key aspects. Firstly, the interaction energy in their system can be completely determined from the position coordinates of the beads, and secondly their functionality is not restricted to one. As a result, there are no additional Boolean degrees of freedom, and complex micellar structures are formed with multiple stickers forming clusters. Within the framework of such a model, for certain parameter values, they observe a first order collapse transition which is driven by competition between the energy gained from forming a core of bridging monomers versus the entropy lost by looping backbone monomers. Another relevant recent work where a first-order transition has been observed is by~\citet{Michieletto2016}, who have examined the 3D dynamics of chromatin folding coupled to 1D dynamics of epigenetic spreading, with a semiflexible bead-spring chain as a model for chromatin fiber. In this model, each bead can have two possible colours, with the colour denoting the epigenetic state and like colours attracting each other. The addition of the colour variable to position coordinates, leads to additional Boolean degrees of freedom, with the interaction energy not being calculable by chain conformation alone. Beads are recoloured periodically with a standard Metropolis acceptance criteria based on the energy difference between beads that are spatially proximate. Unlike in the present model, where the number of stickers is fixed, the number of strongly attracting like-coloured monomers is not constant but calculated dynamically based on proximity and energy of neighbouring monomers. It is observed that a critical value of attraction between like colours exists that separates the chain conformations into a swollen state, with the colours distributed homogeneously along the chain, and a collapsed globular state, with one colour dominant. It is argued that the first-order transition arises because of the coupling between 3D folding dynamics of the polymer and the 1D epigenetic spreading. Interestingly, in contrast to the present model, a second-order transition is never observed for the parameter values that have been examined.

Though the universal swelling of the radius of gyration in dilute sticky polymer solutions in the thermal crossover regime has been demonstrated here with the help of the Soddemann-D\"{u}nweg-Kremer potential, which has many desirable properties, we expect this behaviour to be independent of the specific choice of the excluded volume potential. We hope that this intriguing behaviour predicted by simulations will be tested and validated with careful experiments in the future.

\section*{Conflicts of interest}

There are no conflicts to declare.

\section*{Appendix: The $\theta$-point for homopolymers \label{sec:appendix}} 

\section*{A. The radius of gyration}

\begin{figure}[tbh]
	\begin{center}
			{\includegraphics[width=3.5in,height=!]{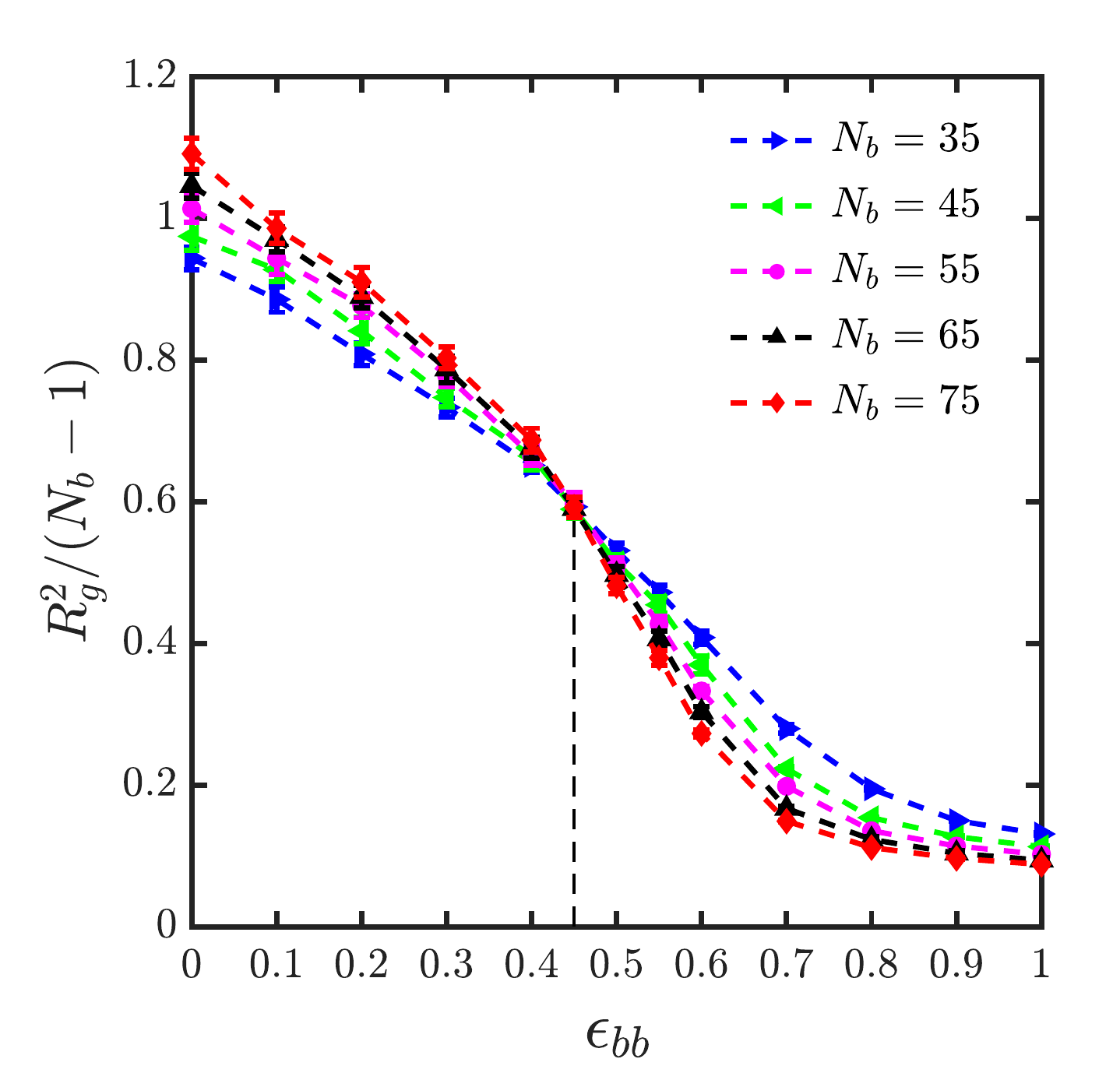}} 
	\end{center}
	\vskip-15pt
	\caption{(Color online) The ratio $R_g^2/(N_b-1)$ versus the well depth of the SDK potential, $\epsilon_{bb}$, to estimate the $\theta$-point for cutoff radius $r_c=1.82\,\sigma$. The symbols represent simulation data and the dotted lines are drawn to guide the eye. The $\theta$-point is estimated as the intersection of all the curves and leads to $\epsilon_{bb}=0.45$.}
	\label{fig:theta_182}
\end{figure}

\begin{figure}[h!]
	\begin{center}
	   \begin{tabular}{c}
		{\includegraphics*[width=3.5in,height=!]{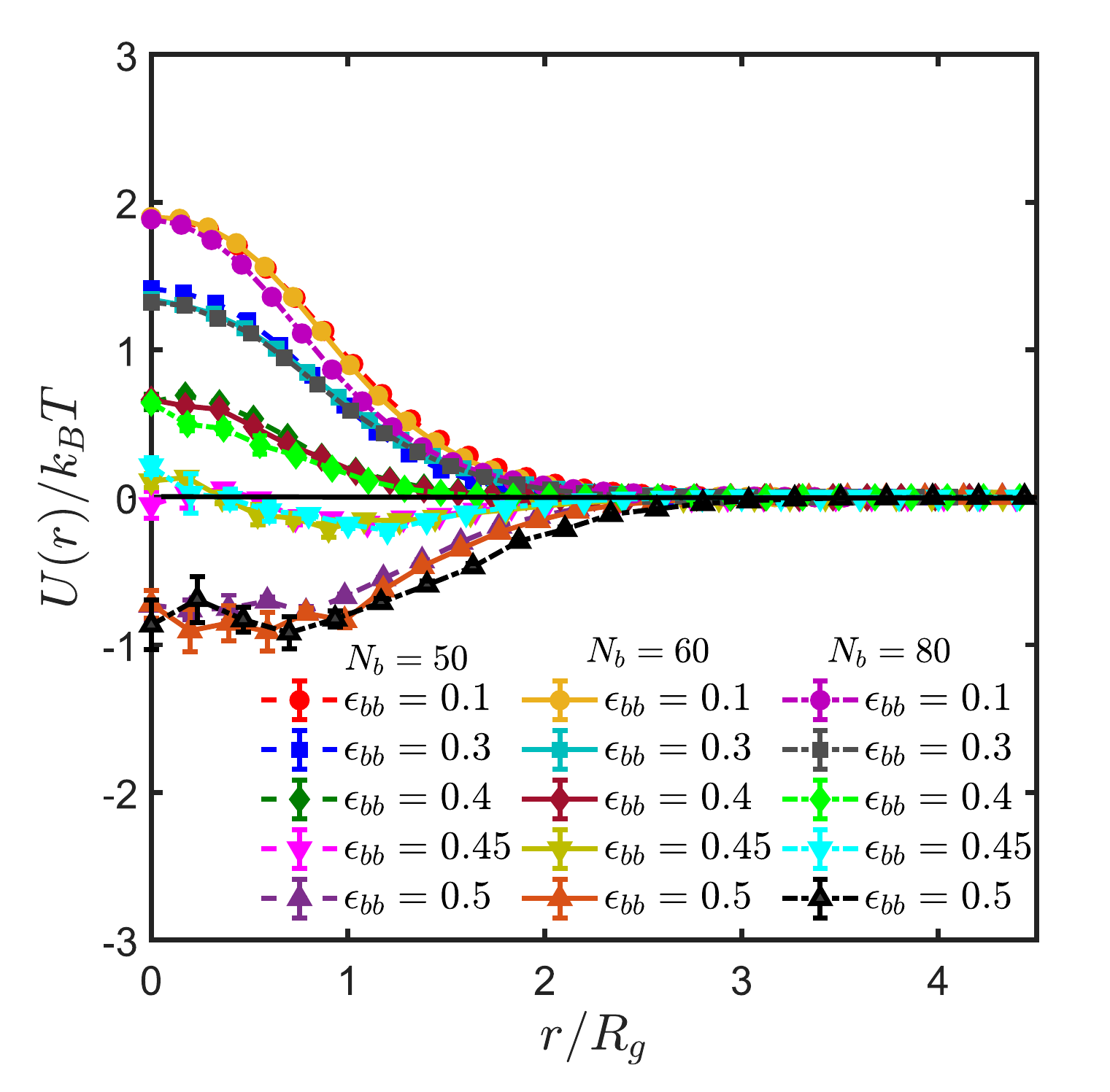}} \\		
		   (a) \\
		{\includegraphics*[width=3.5in,height=!]{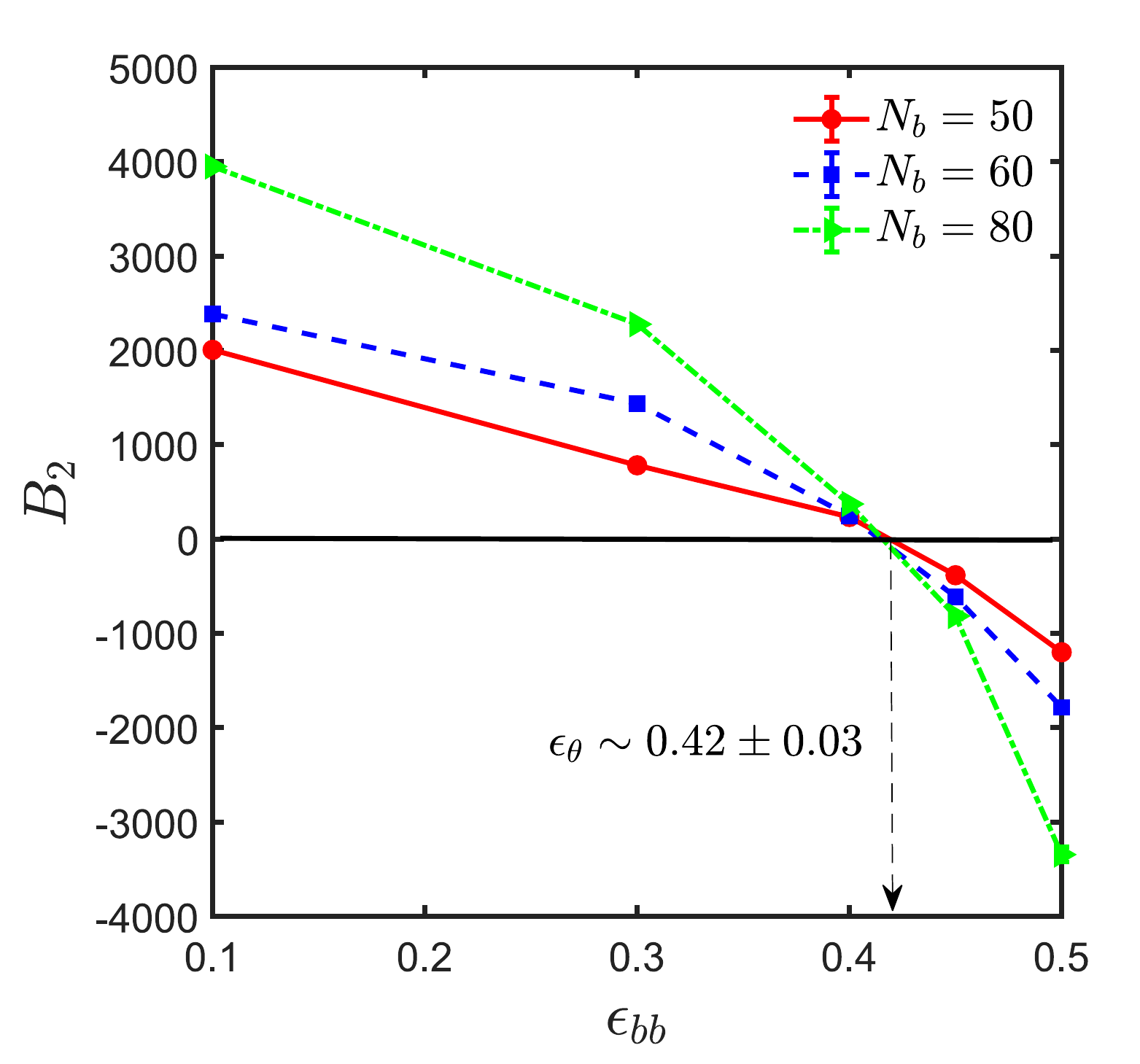}}\\
		   (b) \\
	   \end{tabular}
	\end{center}
	\caption{(Color online) (a) The effective potential, $U(r)/k_BT$, as a function of the separation distance, $r/R_g$, between the centres of mass of a pair of homopolymer chains at different values of potential well depth, $\epsilon_{bb}$. (b) The second osmotic virial coefficient, $B_2$, as a function of potential well-depth, $\epsilon_{bb}$, for different chain lengths. The estimated value of the $\theta$-point, $\epsilon_{\theta}$, from the second virial coefficient is $0.42 \pm 0.03$.}
	\label{fig:homopoly_B2}
\end{figure} 

For a linear polymer chain, the mean-squared radius of gyration follows the universal scaling law $R_g^2 \sim (N_b-1)^{2\nu}$, where the value of the Flory exponent, $\nu$, depends on the solvent quality. At the $\theta$-temperature, linear polymer chains follow RW statistics, with $\nu=1/2$, leading to the ratio $R_g^2/(N_b-1)$ to be independent of the chain length, $N_b$. Whereas, in the case of good and poor solvents, $\nu$ takes the values $3/5$ and $1/3$, respectively\cite{RubCol}. For polymer chains with the SDK potential representing the excluded volume force, the temperature dependence can be captured with the potential well-depth, $\epsilon_{bb}$. As mentioned earlier, $\epsilon_{bb}=0$ (which is equivalent to a WCA potential), represents the athermal limit, where the chain is fully swollen. With increasing values of $\epsilon_{bb}$, a unique value is reached, where the repulsive and attractive interactions between pairs of beads are precisely balanced, leading to $\theta$-like conditions. The value of $\epsilon_{bb}$ at the $\theta$-point can be estimated by plotting the ratio $R_g^2/(N_b-1)$ versus $\epsilon_{bb}$ for different chain lengths, $N_b$, and finding the point of intersection at which curves for different values of $N_b$ intersect, as shown in Fig.~\ref{fig:theta_182}~\cite{Steinhauser,Huissmann2009}. Following this procedure, the $\theta$-point for a homopolymer chain with beads connected by FENE springs having a maximum stretchable length of ${Q_0}^2=50.0$, and $r_c = 1.82 \,\sigma$ as the cut-off radius of the SDK potential, is found to occur at $\epsilon_{bb}=\epsilon_\theta = 0.45$. As discussed in the Supplementary Information, a value of $r_c = 1.5 \,\sigma$ for the cut-off radius leads to the $\theta$-point occurring for a well depth $\epsilon_{bb} = 0.72$, which leads to unphysical asymptotic scaling in the poor solvent limit.

\section*{B. The second virial coefficient}

In addition to the determination of the $\theta$-point from the scaling of radius of gyration, there is an alternative method based on the second osmotic virial coefficient ($B_2$), which involves the determination of the potential of mean force, $U(r)$, between a pair of polymer chains with their centres of mass separated by a distance $r$\cite{Hall1994,Withers2003,Narros2013}. Following the procedure discussed by Dautenhahn et al.\cite{Hall1994}, $U(r)$ is calculated as follows. Two independent chain configurations are chosen from two sets of equilibrated single chain conformations (each having 2000 conformations) and are randomly oriented with respect to each other, with their centres of mass separated by a distance $r$. For a set of values of $r$ ranging from 0 to 5 times the radius of gyration, all such possible two chain configurations (about $4\times 10^6$ configuration pairs) are considered. For each configuration pair $i$, the interaction energy, $\phi_i(r)$, between the two chains is evaluated by computing the pair-wise potential for all pairs of beads, one taken from each chain such that
\begin{equation}
\phi_i(r) = \sum\limits_{p=1}^{N_b}\sum\limits_{q=1}^{N_b}U_{\text{SDK}_i}(r_{pq}).
\end{equation}
Here $\phi_i(r)$ is computed using the SDK potential to account for the pair-wise interaction of the beads and the indices $p$ and $q$ corresponds to chain 1 and 2, respectively. Finally the effective potential is evaluated from $\phi_i(r)$ from the expression 
\begin{equation}\label{Eq:eff_pot}
 \exp{(-U(r)/(k_BT))} = \langle\exp{(-\phi_i(r)/(k_BT))}\rangle
\end{equation}
where the term on the right hand side of Eq.~(\ref{Eq:eff_pot}) is an ensemble average over all configuration pairs. The second virial coefficient is then  easily calculated by evaluating the integral~\cite{RubCol}
\begin{equation}
B_2=\int_0^{\infty} 2 \pi r^2(1-\exp[-U(r)/k_BT]) \,dr.
\label{virial}	
\end{equation}
Positive values of $B_2$ indicate that the polymer solution lies in the good solvent regime, while negative values indicate that it is under poor solvent conditions. At the $\theta$-point, $B_2=0$. 

 Figure~\ref{fig:homopoly_B2} (a) shows the effective potential, $U(r)/k_BT$, as a function of the distance, $r/R_g$, between the centres of mass. For short chains, the potential of mean force depends on chain length, but for sufficiently long chains it saturates within error bars~\cite{Krakoviack,Withers2003,Narros2013}. These are typically fairly small but non-negligible for deep well depths, small distances, and long chains. At small distances sampling is difficult, since for most random pairs the Boltzmann factor is very small due to chain overlap and the strong repulsion of the SDK potential. Therefore, the average is strongly dominated by those few configurations where this is not the case, and this gives rise to an effectively very small sample size~\cite{Krakoviack,Hall1994,Narros2013}. This problem is aggravated for increasing chain lengths and well depths.
 
Figure~\ref{fig:homopoly_B2} (b) is a plot of $B_2$ as a function of the potential well depth, $\epsilon_{bb}$. $B_2$ vanishes at $\epsilon_{bb} = 0.42 \pm 0.03$, independently of chain length. This estimate of the $\theta$-point is in good agreement, within error bars, with the value calculated previously from the scaling of $R_g^2$  ($\epsilon_{\theta} \approx 0.45$). Since both methods lead to approximately the same estimate, the value $\epsilon_{\theta}=0.45$ has been used for all further calculations.

\section*{Acknowledgements}
This research was supported under the Australian Research Council's Discovery Projects funding scheme (project number DP190101825).

 \newcommand{\noop}[1]{}

\end{document}